\documentclass[aps,prl,twocolumn,superscriptaddress,showpacs,floatfix,nobibnotes]{revtex4}
\usepackage{epsfig}
\usepackage{epstopdf}

\usepackage{graphicx}
\usepackage{longtable}
\usepackage{CJK}
\usepackage{color}

\usepackage{mathptmx, courier, pifont}
\usepackage[scaled=0.92]{helvet}
\usepackage[T1]{fontenc}
\usepackage{textcomp}

\begin{document}

\title{Algebraic description of the triaxially to axially rotational shape phase transition}

\author{Yu Zhang }
\affiliation{Department of Physics, Liaoning Normal University,
Dalian 116029, China}

\author{Yu-Xin Liu }
\affiliation{Department of Physics and State Key Laboratory of
Nuclear Physics and Technology, Peking University, Beijing 100871,
China} \affiliation{Center for High Energy Physics, Peking
University, Beijing 100871, China}

\author{Feng Pan}
\affiliation{Department of Physics, Liaoning Normal University,
Dalian 116029, China}\affiliation{Department of Physics and
Astronomy, Louisiana State University, Baton Rouge, LA 70803-4001,
USA}

\author{J. P. Draayer}
\affiliation{Department of Physics and Astronomy, Louisiana State
University, Baton Rouge, LA 70803-4001, USA}

\date{\today}

\begin{abstract}
Within the framework of the interacting boson model, we propose a novel algebraic scheme to describe spin-dependent structural evolutions in triaxial nuclei. Our analysis demonstrates that a triaxially to axially rotational shape phase transition can be induced by the high-order correction which is microscopically motivated by the pairing interaction on the $\gamma$-unstable rotation. This prescription is further applied to describe the structural evolutions along the yrast sequences of $^{132}$Ba and $^{130}$Xe, providing an exemplary case for demonstrating a unified explanation of both low-energy collective structures and high-spin phenomena in O(6)-like nuclei.
\end{abstract}
\pacs{21.10.Re, 21.60.Fw, 27.60.+j}

\maketitle

\begin{center}
\vskip.2cm\textbf{I. Introduction}
\end{center}\vskip.2cm

Quantum phase transitions (QPTs) are recognized as an important way for nuclear system to exhibit shape evolution~\cite{CJC2010,CJ2009,Casten2007,Iachello2011}. QPTs in nuclei typically describe changes in ground-state shapes as a function of nucleon number, hence the terms "shape phase transitions" or "ground-state QPTs". Additionally, phase transitions have been observed to occur as a function of nuclear spin or rotational frequency in individual nucleus. Notable examples include spherical-like nuclei in the $A\sim110$ region, where a transition from vibrational to rotational behavior along the yrast line has been identified~\cite{Regan2003}.
Such spin-driven (or alternatively termed rotation-driven) QPTs provide an alternative explanation of the high-spin backbending behavior~\cite{Regan2003,Cejnar2004} and can be readily identified using the E-GOS curve (E-Gama Over Spin), as highlighted in \cite{Regan2003}. Shape (deformation) transitions along yrast lines can also manifest at very low spins~\cite{Zhang2017}, as observed in the $A\sim150$ region, where some rare-earth nuclei rapidly change their shapes from axial to triaxial ($\gamma$-rigid to $\gamma$-soft) around spins of $4-6\hbar$~\cite{Zhang2021}. Although these transitions often occur gradually compared to backbending phenomena, their signatures remain discernible through characteristic E-GOS curve evolution patterns. These observations collectively demonstrate the significant influence of nuclear rotation on nuclear shapes and associated structural properties.

The interacting boson model (IBM)~\cite{IachelloBook} provides a convenient framework for analyzing spin-driven nuclear transitional phenomena~\cite{Regan2003,Cejnar2004,Liu2006}. Different collective modes in the IBM can be characterized by different dynamical symmetries, which generate the E-GOS curves with varying monotonicity. Changes in monotonicity are used to discern the transitions along the yrast lines of spherical-like nuclei~\cite{Regan2003,Cejnar2004,Liu2006} or axially-deformed nuclei~\cite{Zhang2017}. Despite extensive experimental evidence, including observations in the $A\sim130$ region~\cite{Dewald1988,Paul1989,Higashiyama2002}, a theoretical model specifically addressing spin-driven QPTs in triaxial nuclei remains lacing. In this work, we propose an algebraic approach to describe spin-driven QPT in triaxial nuclei. The resulting model is expected to provide new insights into the transitional phenomena in rotating $\gamma$-unstable (triaxial) systems.

\begin{center}
\vskip.2cm\textbf{II. The Model }
\end{center}\vskip.2cm

In the IBM, three dynamical symmetries (DSs) describe three typical collective modes in nuclei, including the spherical vibrator (U(5)), the axially-deformed rotor (SU(3)) and the $\gamma$-unstable rotor (O(6))~\cite{IachelloBook}. The Hamiltonian in each symmetry limit can be expressed as a combination of the
Casimir operators of the corresponding group chain. A triaxial system in the IBM is conventionally described by the O(6) symmetry limit~\cite{Arima1979}, which exhibits a $\gamma$-unstable picture at the mean-field level~\cite{IachelloBook}. Since the $\gamma$-unstable model and the $\gamma$-rigid model at $\gamma=30^\circ$ (maximally triaxial) may yield similar predictions for most observables~\cite{Zamfir1991}, the terms, "$\gamma$-unstable" and "triaxial", will be used interchangeably without distinction throughout this work.
The O(6) DS in the IBM is characterized by the group
chain
\begin{eqnarray}
\mathrm{U(6)} \supset \mathrm{O(6)} \supset \mathrm{O(5)} \supset
\mathrm{O(3)}\,
\end{eqnarray}
with the O(6) algebra generated by
\begin{eqnarray}\label{Q}
\hat{Q}_u=(d^\dag\tilde{s}+s^\dag\tilde{d})_u^{(2)},~~\hat{T}_u^{(1)}=(d^\dag\tilde{d})_u^{(1)},~\hat{T}_u^{(3)}=(d^\dag\tilde{d})_u^{(3)}\, .
\end{eqnarray}
The corresponding Casimir operators are given by~\cite{IachelloBook}
{\small\begin{eqnarray}\label{casimirO6}
\hat{C}_2[\mathrm{O}(6)]&=&\hat{Q}\cdot\hat{Q}+2\hat{T}^{(1)}\cdot\hat{T}^{(1)}+2\hat{T}^{(3)}\cdot\hat{T}^{(3)}\,
,\\ \label{casimirO5}
\hat{C}_2[\mathrm{O}(5)]&=&2\hat{T}^{(1)}\cdot\hat{T}^{(1)}+2\hat{T}^{(3)}\cdot\hat{T}^{(3)}\,
,\\ \label{casimirO3}
\hat{C}_2[\mathrm{O}(3)]&=&10\hat{T}^{(1)}\cdot\hat{T}^{(1)}\, .
\end{eqnarray}}
A Hamiltonian in the O(6) limit up to two-body interactions can be written as~\cite{Arima1979}
\begin{eqnarray}\label{H6}
\hat{H}_{\mathrm{O}(6)}=A^\prime\hat{C}_2[\mathrm{O}(6)]+B^\prime\hat{C}_2[\mathrm{O}(5)]+C^\prime\hat{C}_2[\mathrm{O}(3)]\,
\end{eqnarray}
with the parameters $A^\prime<0$, $B^\prime\geq0$ and $C^\prime\geq0$.
The energy spectrum can be analytically
expressed as
\begin{equation}\label{EO}
E_\mathrm{O(6)}=A^\prime\sigma(\sigma+4)+B^\prime\tau(\tau+3)+C^\prime
L(L+1)\, ,
\end{equation} where $\sigma$, $\tau$
and $L$ are the quantum numbers used to label the irreducible representations (IRREPs) of O(6), O(5) and O(3). The corresponding eigenstates
are denoted by $|\phi\rangle=|N,\sigma,\tau,\alpha,L,M\rangle$, where $N$ represents the total boson number, $\alpha$ is the
additional quantum number arising from the reduction $\mathrm{O}(5)\supset \mathrm{O}(3)$ with $\alpha=0,~1,\cdots,~[\frac{\tau}{3}]$, and $M$ is the $z$-component of $L$.
From Eq.~(\ref{EO}), it can be deduced~\cite{IachelloBook} that the yrast sequence is generated with $\sigma=N$, $L=2\tau$ and $\tau=0,~1,~2,\cdots,\sigma$. Although
the O(6) Hamiltonian up to two-body terms provides a reasonable description of the low-lying dynamics in triaxial nuclei~\cite{Casten1985},
no transitional phenomena are expected to occur in the boson system as a function of spin.

It is well established~\cite{Berghe1985,Vanthournout1988,Isacker1999,Smirnov2000,Zhang2014,Zhang2022,Zhang2024,Teng2024} that symmetry-conserving high-order terms
can be employed to enrich the dynamical structures of the IBM. For instance,
the three-body interactional term
$(\hat{Q}\times\hat{Q}\times\hat{Q})^{(0)}$ in the O(6) limit may induce a novel axially-rotational
structure~\cite{Isacker1999}, while adding the symmetry-conserving third- and fourth-order terms in the SU(3) limit can realize the
triaxial rotor dynamics~\cite{Smirnov2000,Zhang2014}, by which a theoretical explanation of the anomalous $B(E2)$ in neutron-deficient nuclei has been provided~\cite{Zhang2022,Zhang2024}.
Similarly, symmetry-breaking high-order terms can also used to produce features unattainable with two-body interactions~\cite{Isacker1981,Heyde1984,Sorgunlu2008,Fortunato2011,Zhang2025}.

Here, we reconstruct the O(6) Hamiltonian by introducing the symmetry-conserving high-order term
$(\hat{C}_2[\mathrm{O}(5)])^2$, which preserves an analytical
description of the model structure.
The new model Hamiltonian is designed as follows:
\begin{eqnarray}\label{H}\nonumber
\hat{H}&=&A\hat{C}_2[\mathrm{O}(6)]+B_1(\hat{C}_2[\mathrm{O}(5)])^2+B_2\hat{C}_2[\mathrm{O}(5)]\\
&+&C\hat{C}_2[\mathrm{O}(3)]\,
\end{eqnarray}
with the parameters $A<0,~B_1\leq0,~B_2\geq0,~C\geq0$.
The corresponding energy spectrum can be analytically obtained as
\begin{eqnarray}\label{E}
E=A\sigma(\sigma+4)+B_1\tau^2(\tau+3)^2+B_2\tau(\tau+3)+C L(L+1)\, .
\end{eqnarray}
Compared to Eq.~(\ref{EO}), the newly added O(5) term with $|B_1|\ll |B_2|$ introduces a fourth-order
polynomial in $\tau$, which is expected to be only a minor perturbation
to the $\gamma$-unstable rotation at low spins. However, this term will introduce unexpected new features at high spins that
cannot be generated by an O(6) Hamiltonian limited to two-body interactions. It can be shown that
the model Hamiltonian (\ref{H}) remains well defined for any $N$ if $\mid B_1/B_2\mid<1/(N^2+3N)$, ensuring the ground state IRREP is always determined by $\sigma=N$ and $L=\tau=0$.

Adding the high-order O(5) term can be analogized to the theoretical description of three-dimensional rotation in nuclei.
For an axially-deformed system, the rotational energies are typically expanded in powers of $L(L+1)$ as
\begin{equation}
E_{\mathrm{rot}}=aL(L+1)+bL^2(L+1)^2+\cdots\, ,
\end{equation}
where the high-order terms in $L$ arise from the dependence of the moment of inertia on nuclear spin~\cite{Bohrbook}. This strategy has been dynamically realized in the IBM by Yoshida {\it et al.}~\cite{Yoshida1991} through the introduction of a spin-dependent parameter in the O(3) term, thereby incorporating high-order corrections to the moment of inertia at high spins. Similarly, triaxial rotation in the IBM is described by the O(5) term, and the high-order term in $\tau$ can be analogously thought to come from the expansion of rotational energy in powers of $\tau(\tau+3)$ with
\begin{equation}
E_{\mathrm{trirot}}=a^\prime\tau(\tau+3)+b^\prime\tau^2(\tau+3)^2+\cdots\, .
\end{equation} It will be demonstrated that such high-order corrections can also lead to changes in the moment of inertia at high spins, potentially in a much more dramatic manner.

Apart from the energy expansion, one may question whether there is a priori reason
for introducing these high-order terms from a microscopic perspective. Previous analysis~\cite{Yoshida1991} has indicated that the high-order O(3) terms affecting the moment of inertia can be microscopically explained from the pairing correlation of nucleons.
If the single-$j$ shell approximation is adopted, the pairing interaction can be written as~\cite{Otsuka1978}
\begin{eqnarray}
V=-G_\nu S_\nu^\dag S_\nu-G_\pi S_\pi^\dag S_\pi\, ,
\end{eqnarray}
where
\begin{equation}
S_\rho^\dag=\frac{1}{2}(-1)^{j_\rho-m_\rho}a_{j_\rho m_\rho}^\dag a_{\rho j_\rho-m_\rho}^\dag,~~\rho=\nu,~\pi\,
\end{equation}
with $a_{jm}^\dag$ representing the nucleon creation operator.  The IBM image of the pairing interaction has been derived in \cite{Yoshida1991} through the OAI mapping~\cite{Otsuka1978} and the projection technique~\cite{Harter1985}. It is given, apart from a constant, by
\begin{eqnarray}\label{VIBM}
V^{\mathrm{IBM}}=y_1 \hat{n}_d+y_2 \hat{n}_d^2\, ,
\end{eqnarray}
where $\hat{n}_d=\sum_md_m^\dag d_m$.
The corresponding parameters were determined as~\cite{Yoshida1991}
\begin{eqnarray}
&&y_1=\Big(\frac{N_\nu}{N}G_\nu\Omega_\nu+\frac{N_\pi}{N}G_\pi\Omega_\pi\Big)-y_2,\\
&&y_2=-\Big(\frac{N_\nu(N_\nu-1)}{N(N-1)}G_\nu+\frac{N_\pi(N_\pi-1)}{N(N-1)}G_\pi\Big)\, ,
\end{eqnarray}
where $\Omega_{\pi(\nu)}=j_{\pi(\nu)}+\frac{1}{2}$ and $N_\pi(N_\nu)$
denotes the number of the proton (neutron) bosons with $N=N_\pi+N_\nu$.

With the IBM image of the pairing interaction, one can examine how the pairing interaction influences different modes in the IBM framework. Specifically, we have worked out the expectations of $V^{\mathrm{IBM}}$ defined in (\ref{VIBM}) under the wave functions for the ground-state bands in the U(5), SU(3), and O(6) symmetry limits using the Wigner coefficients tabulated in \cite{Vergados1968,Isacker1984}. The final results can be analytically expressed as follows:
\begin{eqnarray}\label{u5}
&&\langle V^{\mathrm{IBM}}\rangle_{\mathrm{U(5)}}=y_1n_d+y_2n_d^2,\\ \label{su3}
&&\langle V^{\mathrm{IBM}}\rangle_{\mathrm{SU(3)}}=z_1L(L+1)+z_2L^2(L+1)^2+z_0,\\ \label{o6}
&&\langle V^{\mathrm{IBM}}\rangle_{\mathrm{O(6)}}=w_1\tau(\tau+3)+w_2\tau^2(\tau+3)^2+w_0\, ,
\end{eqnarray}
in which the corresponding parameters are obtained as
\begin{eqnarray}
&&z_1=y_1\frac{1}{6(2N-1)}+y_2\frac{4N^2-11N+4}{9(2N-3)(2N-1))},\\
&&z_2=y_2\frac{1}{36(2N-3)(2N-1)},\\
&&z_0=y_1\frac{4(N^2-N)}{3(2N-1)}+y_2\frac{8N(N-1)^2}{9(2N-3)}\,
\end{eqnarray}
and
\begin{eqnarray}
&&w_1=y_1\frac{1}{2(N+1)}+y_2\frac{N^2-3N+1}{2(N^2+N)},\\
&&w_2=y_2\frac{1}{4(N^2+N)},\\
&&w_0=y_1\frac{N^2-N}{2(N+1)}+y_2\frac{N^3-2N^2+7N-6}{4(N+1)}\, .
\end{eqnarray}
It is evident that the IBM image of the pairing interaction indeed introduces a second-order correction to the ground-state band in each symmetry limit. Specifically,
the results indicate that the strengths of the second-order terms are all negative and significantly smaller compared to
the leading-order terms. For instance, it is given by $\mid w_2/w_1\mid\sim1/N^2$, which aligns precisely
with the parameter constraints on the model Hamiltonian (\ref{H}) discussed above. Moreover, the negative contribution from the high-order terms suggests that the effects of the pairing interaction on rotational spectrum may diminish as a function of nuclear spin. These observations are consistent with the traditional understanding of pairing correlation in nuclei~\cite{Mottelson1960} and previous analyses of the relevant topics based on the IBM~\cite{Yoshida1991,Pan1992}. In short, the high-order O(5) term appearing in the new model can be microscopically justified by the pairing interaction.
Note that expressions for pairing motivated high-order terms similar to those in Eq.~(\ref{su3})-(\ref{o6}) were also provided in \cite{Yoshida1991,Pan1992}. Additionally, a negative $n_d^2$ term like that given in Eq.~(\ref{u5}) has been shown to play a crucial role in modeling the collective backbending~\cite{Long1997} and rotation-driven QPT in spherical-like nuclei~\cite{Liu2006}. Consequently, it appears that the high-order terms given in (\ref{u5})-(\ref{o6}) are generally necessary to simulate the pairing effects at high spins within the IBM framework.

\begin{figure}
\begin{center}
\includegraphics[scale=0.3]{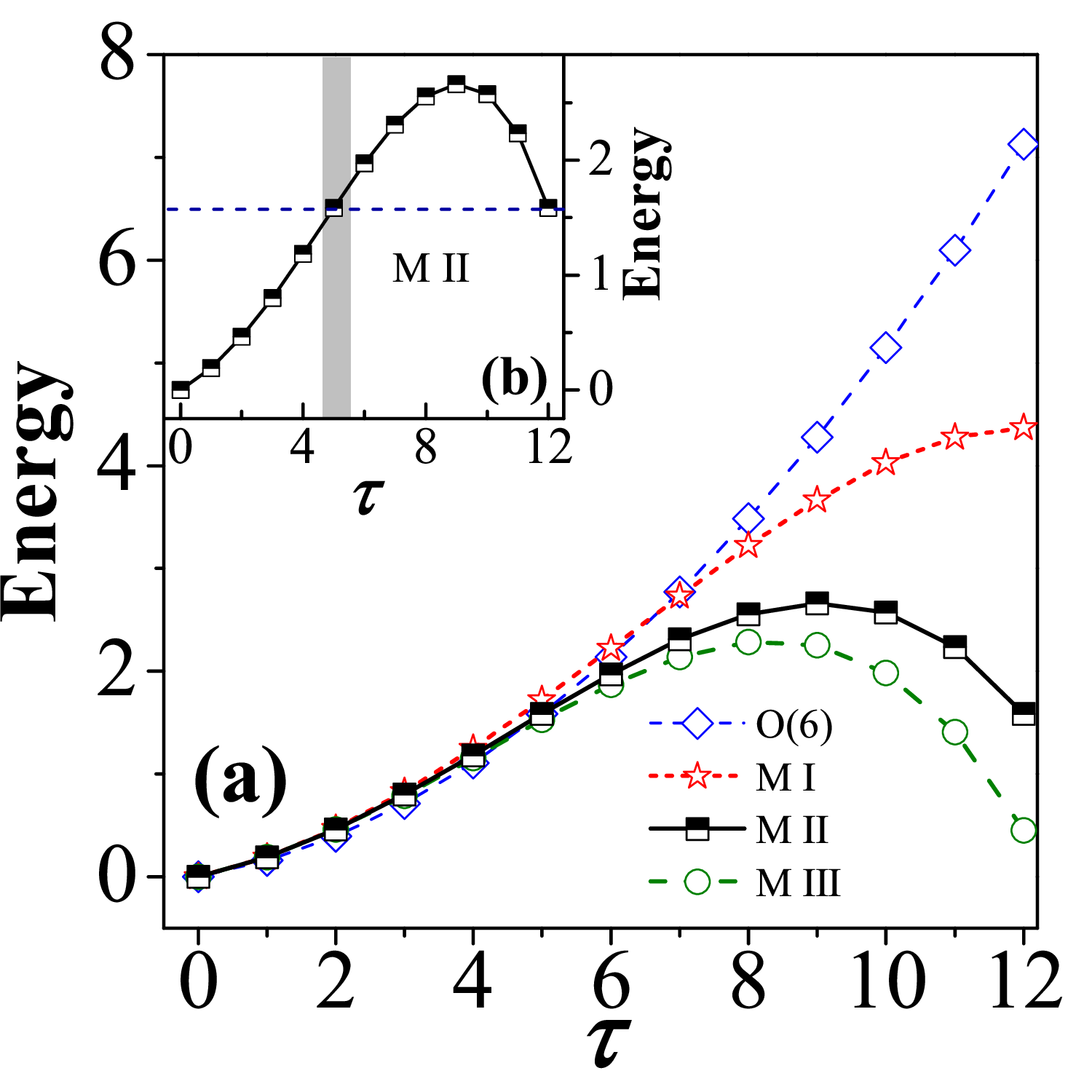}
\caption{The evolution of energy (in any units) against the seniority number $\tau$ is shown for states contained in the O(6) IRREP with $\sigma=N$. The results are obtained from the new model Hamiltonian (\ref{H}) and the traditional O(6) Hamiltonian (\ref{H6}). In the calculations, the boson number is set to $N=12$, and three different values of $B_1$ are selected to examine the effect of the high-order term: $B_1=-0.000134$ (M I), $B_1=-0.00022$ (M II) and $B_1=-0.000255$ (M III). The other parameters are fixed at $A=A^\prime=-0.01$, $B_2=0.0484$, $B^\prime=0.0396$ and $C=C^\prime=0$. The inset highlights the results obtained from the new model (M II). \label{F1}}
\end{center}
\end{figure}

\begin{figure*}
\begin{center}
\includegraphics[scale=0.23]{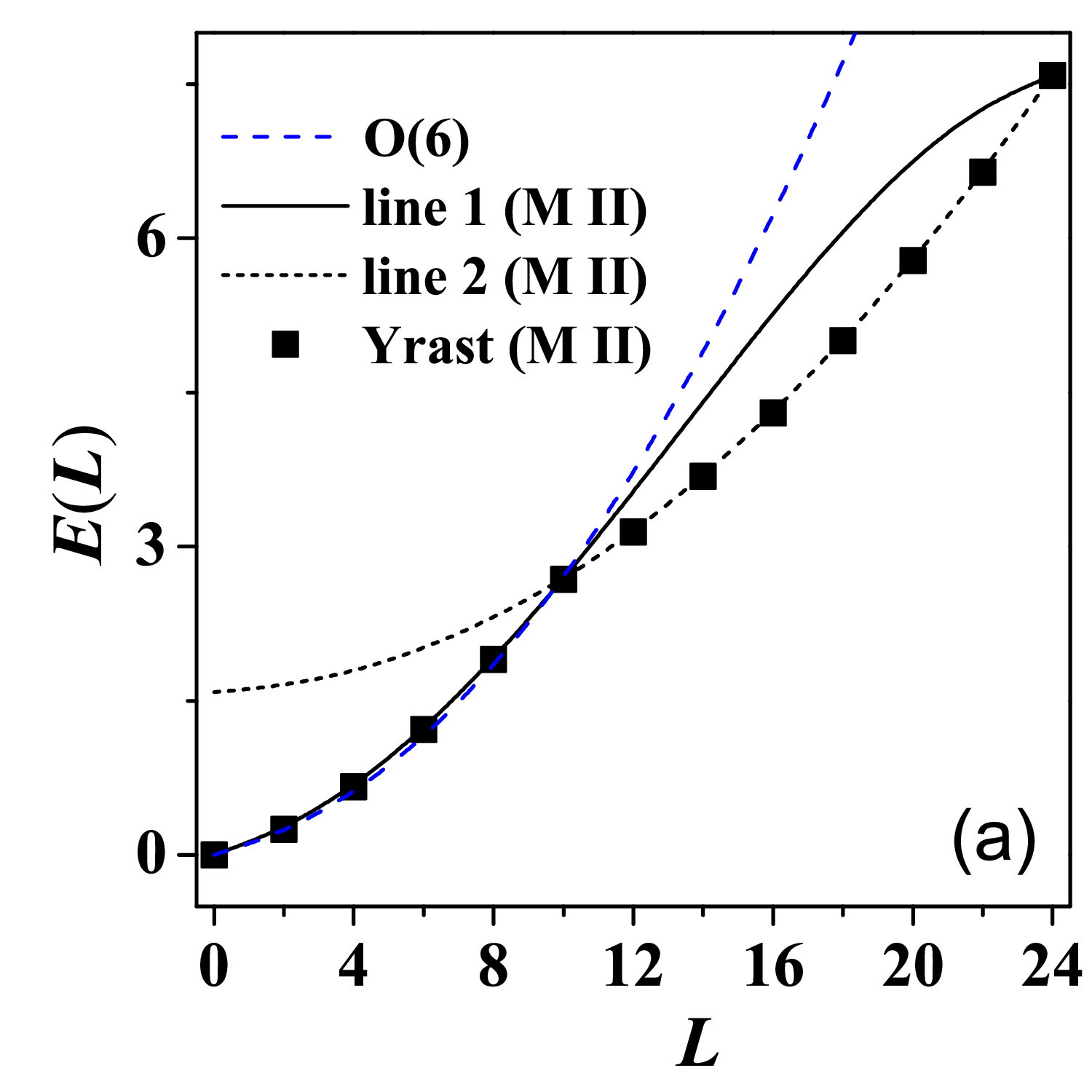}
\includegraphics[scale=0.23]{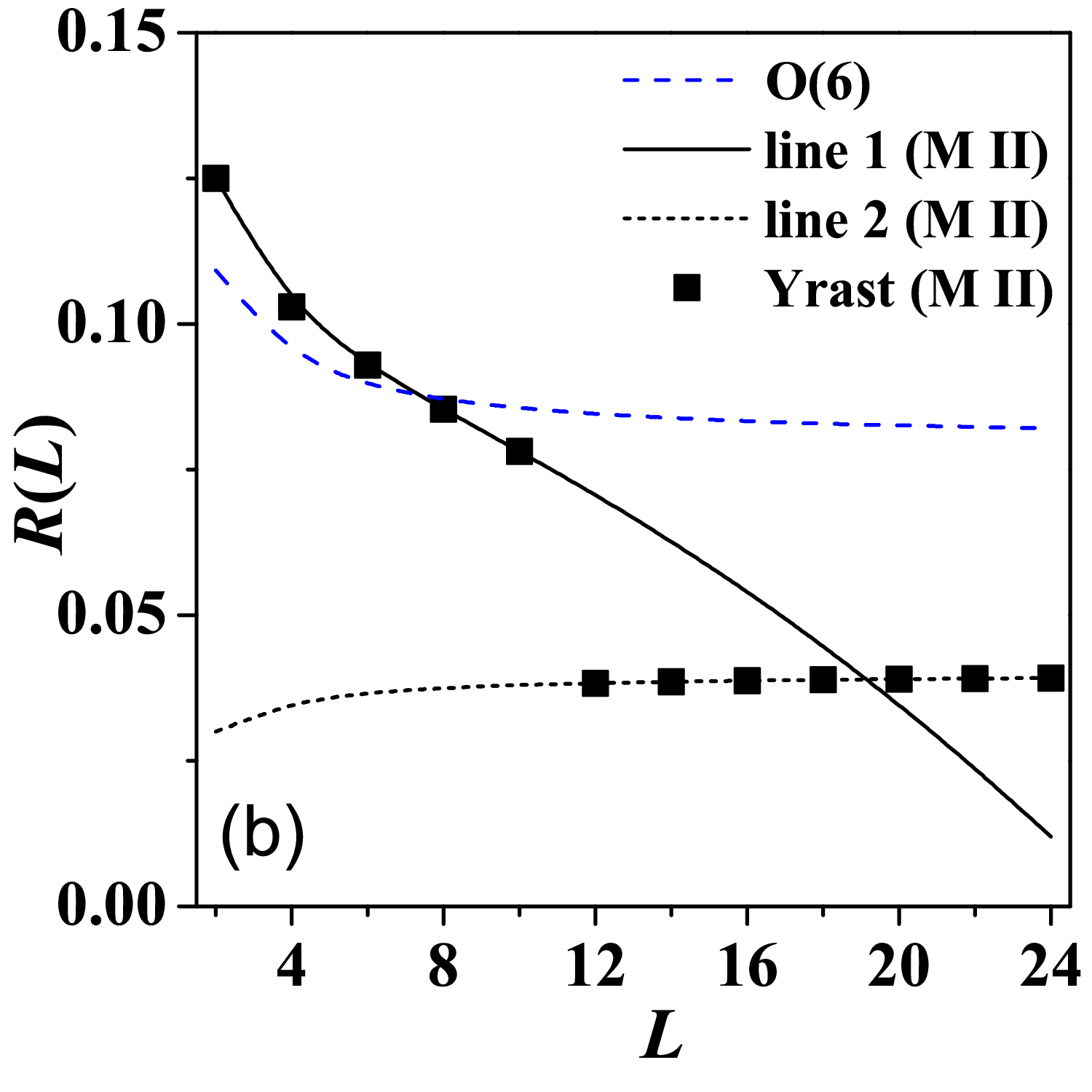}
\includegraphics[scale=0.23]{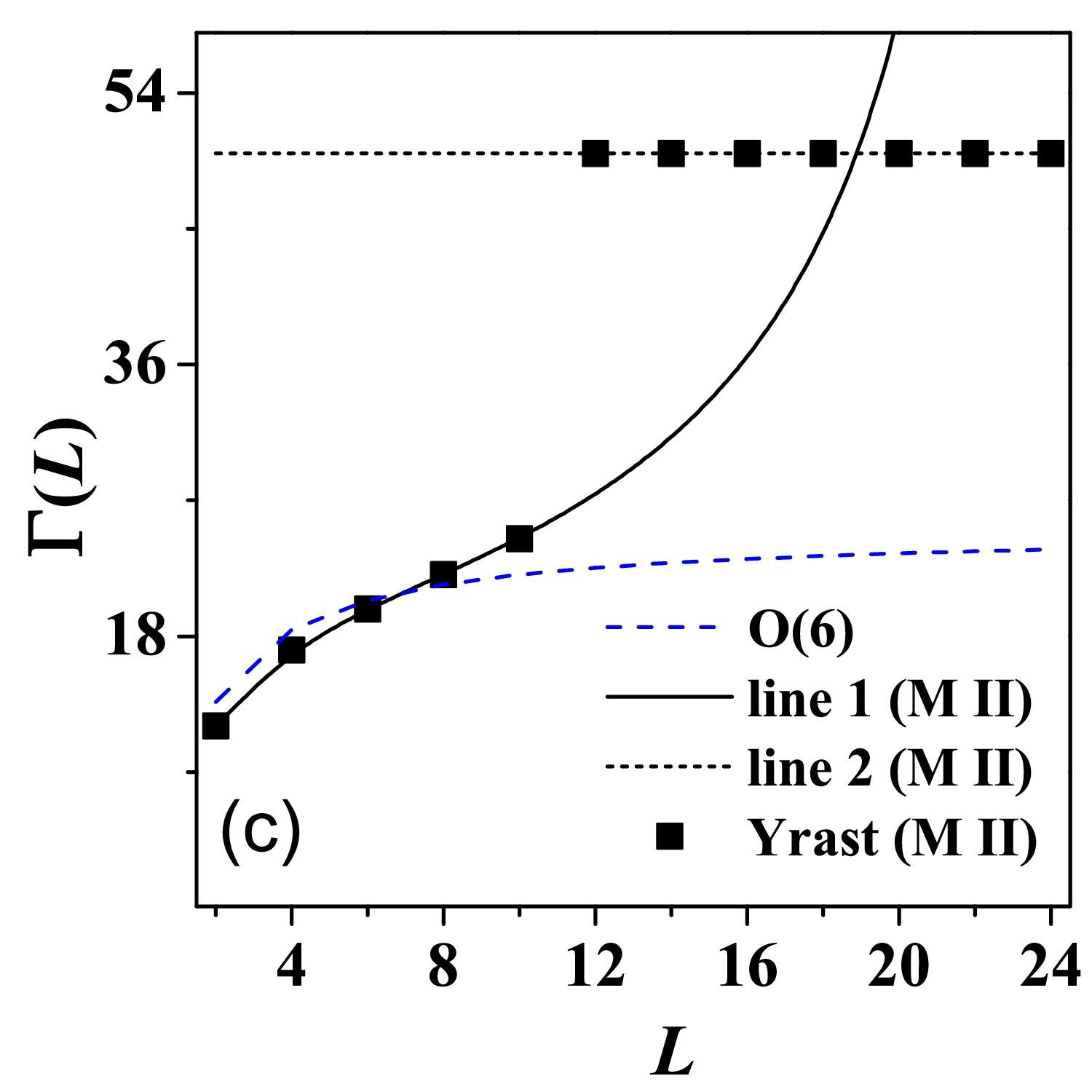}
\caption{(a) The level energies (in any units) evolve as a function of $L$. Line 1 and line 2 represent
the new model results for $\tau=L/2$ and $\tau=N$, respectively, while the O(6) line denotes the results extracted from the traditional O(6) description.
(b) The same as in (a), but for the E-GOS curves described by $R(L)$. (c) The same as in (a), but for the
effective moments of inertia, $\Gamma(L)$. In the calculations, the parameters for both the new model (M II) and the O(6) Hamiltonian are the same as those used in FIG.~\ref{F1}, except for $C=C^\prime=0.01$. \label{F2}}
\end{center}
\end{figure*}

\begin{center}
\vskip.2cm\textbf{III. Rotational QPT along the
Yast Line}
\end{center}\vskip.2cm

\begin{center}
\textbf{(A)~\it{Analytical analysis}}
\end{center}\vskip.2cm

To demonstrate the features of the new model, the energy spectra for the lowest O(6) IRREP $\sigma=N$ solved from Eq.~(\ref{H}) are compared with those derived from Eq.~(\ref{H6}).
As shown in FIG.~\ref{F1}, the level energies in both cases are nearly identical within the range $\tau\in[0,~5]$, suggesting
that the high-order term in the new model introduces only a minor perturbation to the low-lying O(6) spectrum. However, significant discrepancies appear for
$\tau\in[5,~N]$. In the traditional O(6)
description, the level energies  increase monotonically with $\tau$, whereas the new model exhibits a parabola trend over the entire range $\tau\in[0,~N]$ in the M II or M III scenarios. The inset further
highlights the existence of a critical $\tau$ value, which is determined by the energy relation
\begin{equation}
E_{\tau_\mathrm{c}}\leq
E_{\tau_\mathrm{max}}<E_{\tau_\mathrm{c}+1}\, ,
\end{equation}
where $\tau_\mathrm{max}$ denotes the maximal value of $\tau$.
In other words, once $\tau>\tau_c$, the configuration with $\tau=\tau_\mathrm{max}$ becomes energetically favorable.
For example, the critical value in the M II case is given by
$\tau_\mathrm{c}=5$, in contrast to $\tau_\mathrm{max}=12$.
More generally, for $\sigma=N$, the critical $\tau$ value is given by
\begin{equation}\label{tc}
\tau_\mathrm{c}=-3/2+\sqrt{9-4(B_2/B_1+N^2+3N)}/2\,
\end{equation}
with the critical angular momentum $L_\mathrm{c}=2\tau_\mathrm{c}$.
Given that $0\leq\tau_c\leq N$, it is further required that
\begin{eqnarray}
1/(2N^2+6N)\leq\mid B_1/B_2\mid\leq1/(N^2+3N)\, ,
\end{eqnarray} indicating that the strength of the high-order term is at most of $1/N^2$ order
relative to the two-body term. If $\mid B_1/B_2\mid<1/(2N^2+6N)$, as in the M I case shown in FIG.~\ref{F1}, there would be minimal differences between
the new model and the traditional O(6) description, except for an energy correction on the $\tau(\tau+3)$ rule at high spins.
Such an energy correction, referred to as $\tau$-compression, was previously analyzed by Pan {\it et~al.}~\cite{Pan1992} through a pairing-motivated boson interaction.
In this work, we will focus on the more
interesting scenario where $\mid B_1/B_2\mid\geq1/(2N^2+6N)$, exemplified by the M II case in FIG.~\ref{F1}. To this end,
we will use the M II case to illustrate the evolution of the yrast state in the new model.

Yrast state is defined as the lowest-energy state for a given nuclear spin. Any change in the properties of the yrast states
may indicate an alteration in the collective mode of a rotating system.
In the new model, the yrast state configurations for
$L\leq L_\mathrm{c}$ in the M II case are distinctly different from those for
$L>L_\mathrm{c}$. According to the branching rules of O(6) IRREPs~\cite{IachelloBook},
the configurations for $L\leq L_\mathrm{c}$ are characterized by
$|N,\sigma=N,\tau,\alpha=0,L=2\tau,M\rangle$ with
$\tau=0,~1,~2,~\cdots,~\tau_\mathrm{c}$, while those for $L>L_\mathrm{c}$
transition to $|N,\sigma=N,\tau=N,\alpha=0,L,M\rangle$ with
$L=2\tau_\mathrm{c}+2,~2\tau_\mathrm{c}+4,~\cdots,~2N$, corresponding to the $d$-boson condensate states. Consequently, it can be concluded that a spin-driven QPT may occur
along the yrast line in the new model. This transitional behavior is a manifestation of the dynamical symmetry breaking of O(5)$\supset$O(3). The related
discussions on $\sigma=N$ can be directly generalized to
$\sigma=N-2,~N-4,\cdots$ with the critical $\tau$ value given by
\begin{equation}
\tau_\mathrm{c}=-3/2+\sqrt{9-4(B_2/B_1+\sigma^2+3\sigma)}/2\, .
\end{equation}

To identify the spin-driven QPT features, the yrast level energies $E(L)$
and the associated E-GOS curve described by
$R(L)=\frac{E_\gamma(L\rightarrow L-2)}{L}$~\cite{Regan2003} are calculated for the new model (the M II case).
As noted above, the E-GOS curve is a simple yet powerful tool to discern collective modes in the IBM. It follows
$R(L)\propto\frac{1}{L}$ for the spherical vibrator (U(5) limit), $R(L)\propto(4-\frac{1}{L})$ for the axially-deformed rotor (SU(3) limit) and $R(L)\propto(1+\frac{2}{L})$ for the triaxial rotor (O(6) limit)~\cite{Zhang2017}. Additionally, the effective moment of inertia defined by $\Gamma(L)=\frac{2L-1}{E_\gamma(L\rightarrow L-2)}$ is also considered, as different collective modes may generate distinct types of moments of inertia~\cite{Bohrbook}.
All the calculated results are presented as a function of $L$ and shown in FIG.~\ref{F2}, where quantities derived from the traditional O(6) Hamiltonian (\ref{H6}) are provided for comparison.
As seen from FIG.~\ref{F2}(a), the level energies in the new model can be divided into two serials based on their collective
configurations: one (line 1) is characterized by
$|N,\sigma=N,\tau,\alpha=0,L=2\tau,M\rangle$ and the other (line
2) is described by $|N,\sigma=N,\tau=N,\alpha=0,L,M\rangle$.
The yrast states in the new model initially evolve along
line 1 up to $L=10$ and then turn to follow line 2 untill $L=2N$. Line 1 describes triaxial rotation with
configurations being identical to those obtained from the traditional O(6) Hamiltonian, while line 2 represents axial rotation complying with the $L(L+1)$ law. The
sudden change in the yrast state configurations is identified here as a triaxial-to-axial rotational QPT. This transition is further highlighted by
the E-GOS curves shown in FIG.~\ref{F2}(b). It is shown that $R(L)$ may suddenly change around $L_\mathrm{c}=10$ from a decreasing trend along line 1 to a slight
increase along line 2, indicating the shift from triaxial to axial mode~\cite{Zhang2017}. Likewise, the transitional process is vividly demonstrated
by the evolution of $\Gamma(L)$, as shown in FIG.~\ref{F2}(c). One can observe that
$\Gamma(L)$ is $L$-dependent in the triaxial mode and then abruptly transitions to a constant in the axial mode.

\begin{figure}
\begin{center}
\includegraphics[scale=0.35]{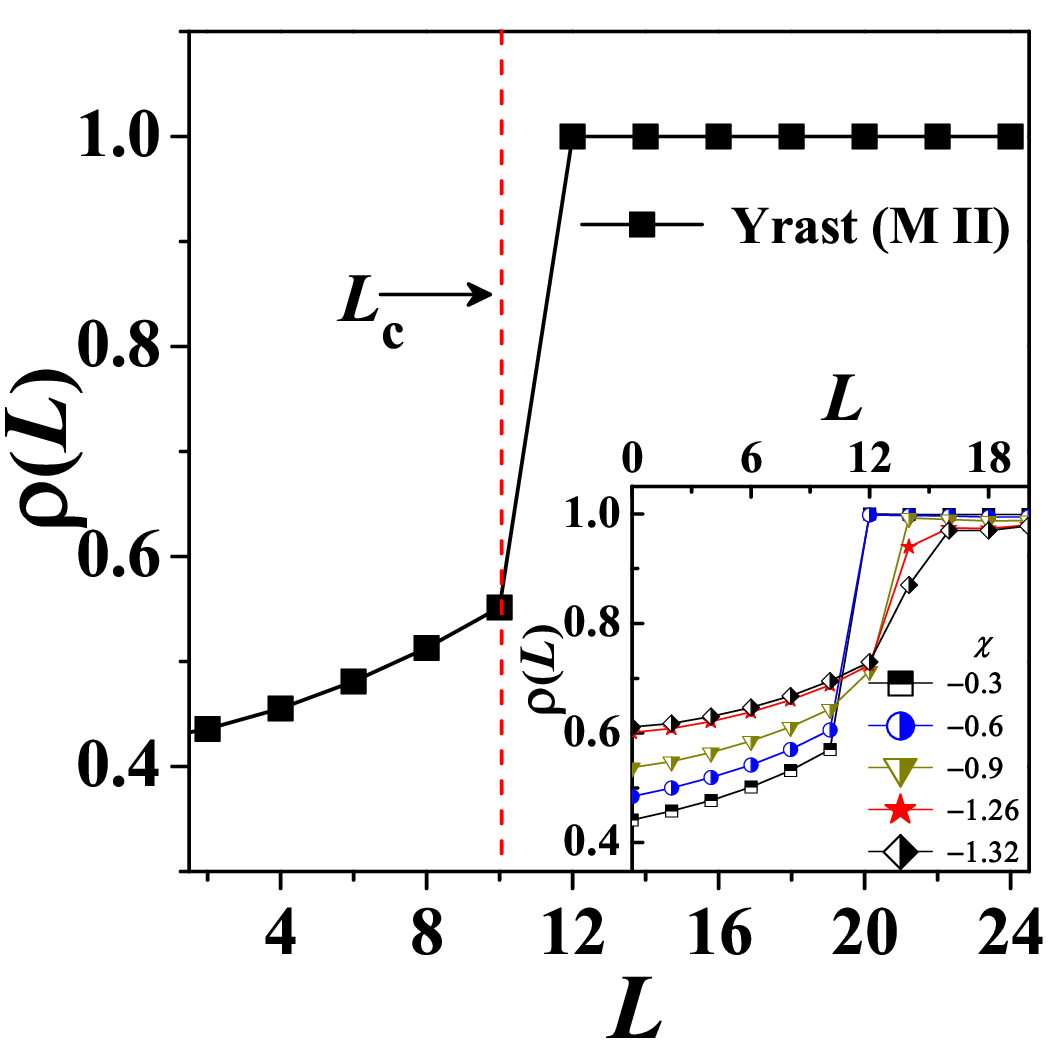}
\caption{The $d$-boson occupation probability $\rho(L)$ for the yrast states is shown. In the calculation, the parameters are taken the same as those in FIG.~\ref{F2},
namely $A=-0.01$, $B_1=-0.00022$, $B_2=0.0484$ and $C=0.01$. The inset highlights the results for the symmetry-breaking cases with $\chi\neq0$ (see the text),
in contrast to the MII case corresponding to $\chi=0$.\label{F3}}
\end{center}
\end{figure}

Apart from the excitation energies, some critical features can be identified from the $B(E2)$ transitions. In the
IBM, the $E2$ transitional operator is chosen as~\cite{IachelloBook}
\begin{equation}\label{E2}
T_u^{E2}=e_\alpha\Big[[d^\dag\times\tilde{s}+s^\dag\times\tilde{d}]_u^{(2)}+\theta[d^\dag\times\tilde{d}]_u^{(2)}\Big]\,
,
\end{equation}
where $e_\alpha$ represents the effective charge and $\theta$ denotes a dimensionless parameter. Since the model
is designed in the O(6) limit, one can obtain useful insights without performing detailed calculations~\cite{IachelloBook}.
The first term in (\ref{E2}) follows the selection rule $\Delta\tau=\pm1$ and the second term follows the
rule $\Delta\tau=0,\pm2$. This implies that for yrast states with
$L_1\leq L_\mathrm{c}$, the $B(E2)$ transitions $B(E2;L_1\rightarrow L_1-2)$ are determined solely by the first term, whereas for $L_1>L_\mathrm{c}$, they are induced by the second term.
Consequently, the transition
$B(E2;L_\mathrm{c}+2\rightarrow L_\mathrm{c})$ is completely prohibited due to $\Delta\tau>2$.
Therefore, a much weaker experimental transition for $B(E2;L_\mathrm{c}+2\rightarrow L_\mathrm{c})$ could serve as a signature of the spin-drive QPT.
Additionally, nonzero values of the quadrupole moments, defined as $Q(L)=\langle L~ M=L|\sqrt{\frac{16\pi}{5}}T_0^{E2}|L~ M=L\rangle$, will be achieved in the O(6) limit if $\theta\neq0$
is considered in (\ref{E2}).

\begin{center}
\textbf{(B)~\it{Order parameter and Cranking analysis}}
\end{center}\vskip.2cm

To better understand this spin-driven QPT, the $d$-boson occupation probability
defined by $\rho(L)=\langle L_1|\hat{n}_d|L_1\rangle/N$ is examined, and the results are presented as a function of $L$ in FIG.~\ref{F3}.
The $d$-boson occupation probability has been considered a quantum order parameter for the ground-state QPTs in nuclei~\cite{Iachello2004}. Here, it is supposed that this quantity can serve a similar role of
the spin-driven QPT. The difference lies in the control parameter: while nucleon number serves as the control parameter in the former~\cite{Casten2007}. nuclear spin takes on this role in the latter~\cite{Regan2003}. As shown in FIG.~\ref{F3}, the order parameter increase gradually up to
$\rho\sim0.5$ and then abruptly jumps to $\rho=1.0$ for $L>10$. This discontinuous change in the order parameter is characteristic of a 1st-order ground-state QPT~\cite{Iachello2004}. Consequently, the current spin-driven transition can be classified as a 1st-order QPT. In fact, discontinuous changes have already been observed in the evolutions of $R(L)$ and $\Gamma(L)$ as discussed above, indicating that these quantities can also act effective order parameter~\cite{Iachello2004} to detect the spin-driven QPT experimentally. It should be emphasized that discontinuities in order parameters primarily arise from changes in wave functions (or collective configurations) rather than the discrete values of nuclear spin or nucleon number, reflecting the intrinsic nature of QPTs in finite nuclear systems~\cite{Casten1999}.

Undoubtedly, the model designed in the O(6) limit can be extended to more general case without conserving the O(6) DS. An interesting question is to what extent the QPT features
can persist in an O(6) DS breaking scenario. One way to break the O(6) DS is by adding a U(5) term
to the model Hamiltonian, which may still preserve the O(5) DS.
Here, we aim to explore a more general situation that breaks the DS at both the O(6) and O(5) levels.
To achieve this, we replace the quadrupole operator $\hat{Q}$ defined in (\ref{Q})
with a more general form~\cite{Warner1983}
\begin{eqnarray}\label{Qx}
\hat{Q}_u^\chi=(d^\dag\tilde{s}+s^\dag\tilde{d})_u^{(2)}+\chi[d^\dag\times\tilde{d}]_u^{(2)}\, ,
\end{eqnarray} where the parameter varies with $\chi\in[0,-\sqrt{7}/2]$. The resulting Hamiltonian is now written as
\begin{eqnarray}\label{Hp}
\hat{H}^\prime=A\hat{H}_\chi+B_1(\hat{C}_2[\mathrm{O}(5)])^2+B_2\hat{C}_2[\mathrm{O}(5)]+C\hat{C}_2[\mathrm{O}(3)]\,
\end{eqnarray}
with
\begin{eqnarray}
\hat{H}_\chi=\hat{Q}^\chi\cdot\hat{Q}^\chi+\hat{C}_2[\mathrm{O}(5)]\, .
\end{eqnarray}
Then, using Eq.~(\ref{Q})-Eq.~(\ref{casimirO5}), one can derive that
\begin{eqnarray}
\hat{H}_{\chi=0}=\hat{C}_2[\mathrm{O}(6)]\, .
\end{eqnarray}
This demonstrate that the Hamiltonian (\ref{Hp}) at $\chi=0$
reduces exactly to the model Hamiltonian (\ref{H}), indicating that the parameter $\chi$ serves as a measure of deviation from the O(6) DS. In the inset of FIG.~\ref{F3}, we examine the evolution of the order parameter $\rho(L)$, solved from the Hamiltonian (\ref{Hp}) for various values of $\chi$. It is evident that discontinuous changes in $\rho(L)$ are clearly observed for $\chi\neq0$. This suggests that the QPT features are well preserved even when the system deviates from the O(6) DS, although the critical points shift from $L_\mathrm{c}=10$ to $L_\mathrm{c}=12$ when $|\chi|>0.6$.

\begin{figure}
\begin{center}
\includegraphics[scale=0.165]{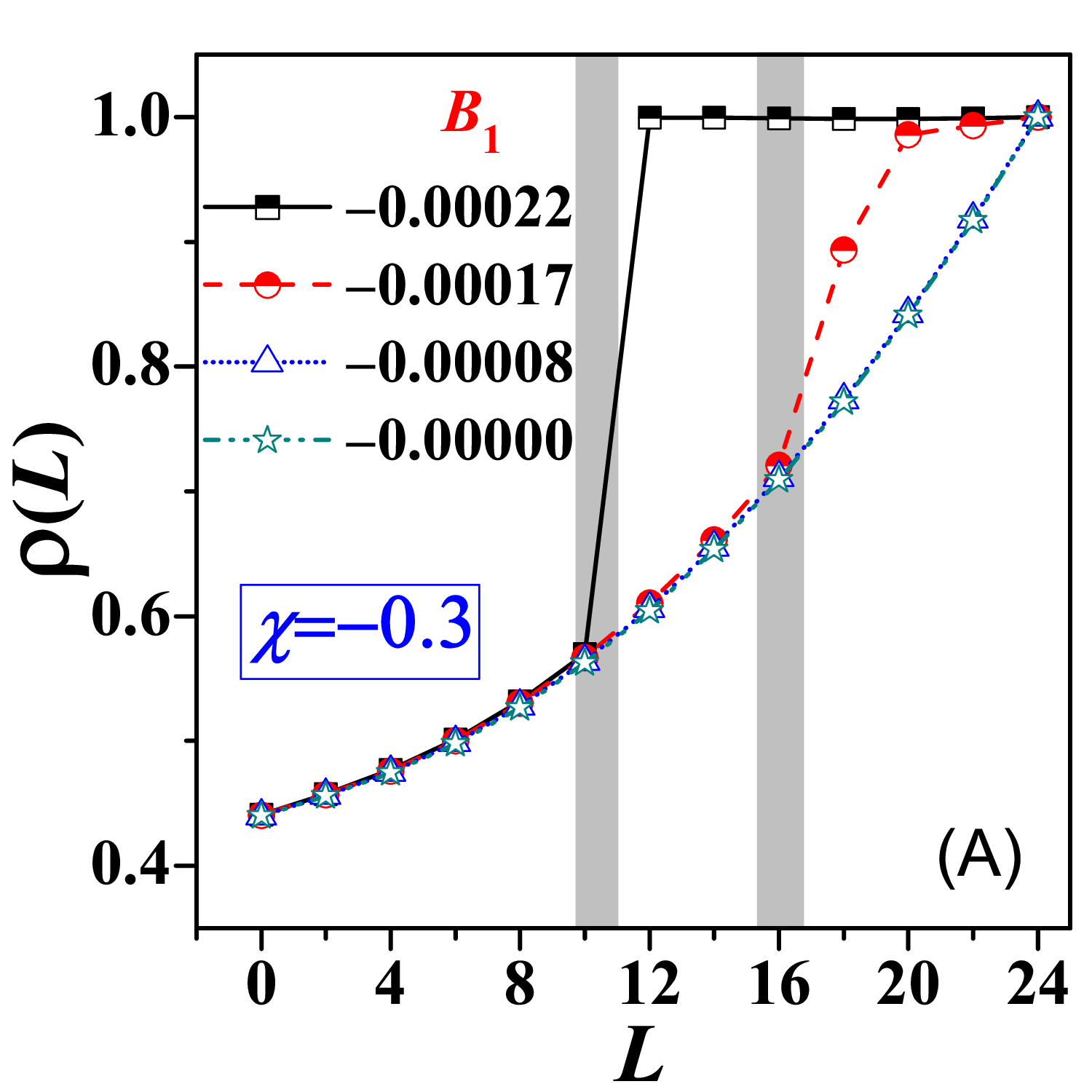}
\includegraphics[scale=0.165]{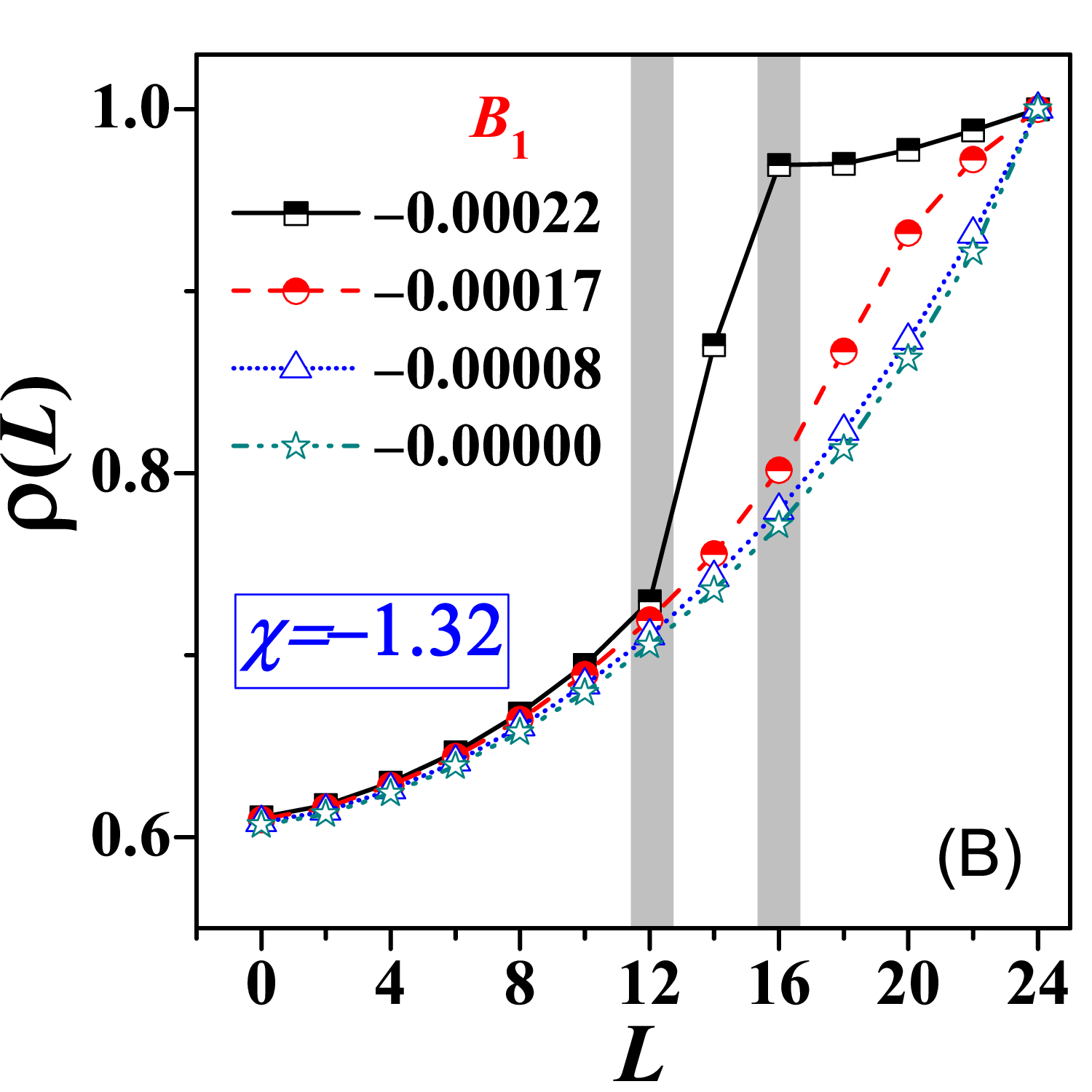}
\includegraphics[scale=0.165]{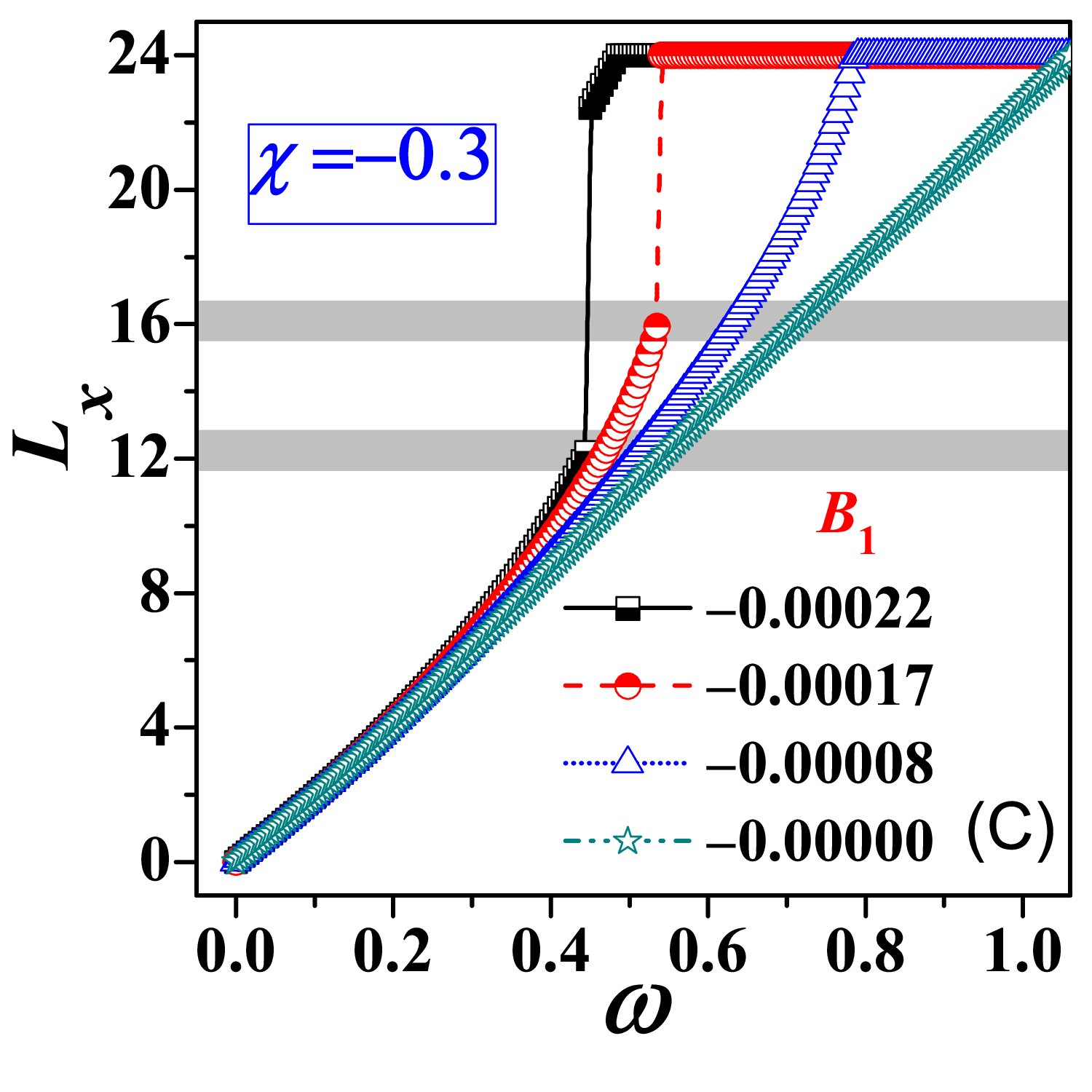}
\includegraphics[scale=0.165]{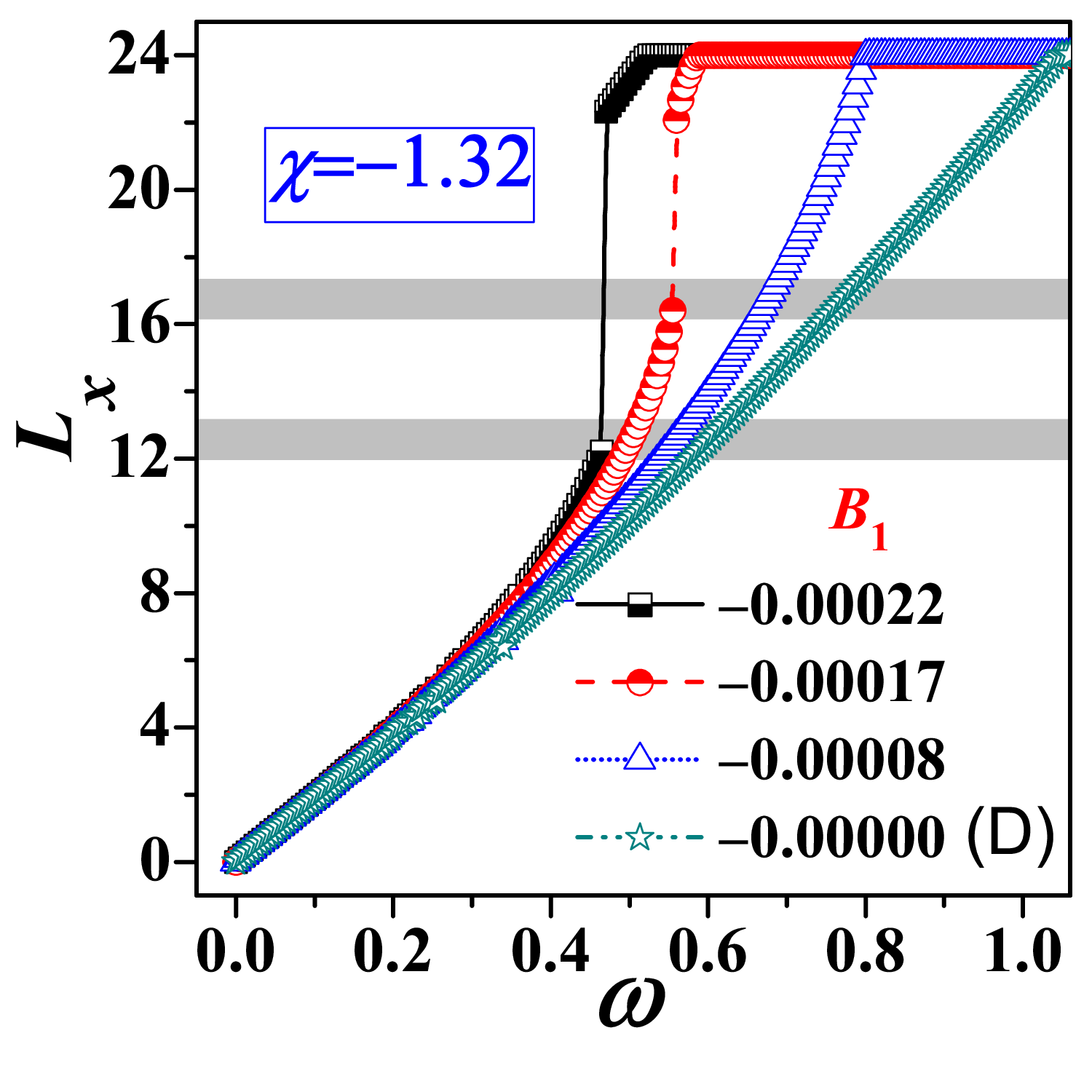}
\caption{(A) The $\rho(L)$ evolutions for the case with $\chi=-0.3$, as shown in the inset of FIG.~\ref{F3}, but for different $B_1$ values.
(B) The same as in (A), but for the case with $\chi=-1.32$. (C) The cranking results for $L_x$ vs $\omega$, corresponding to the cases in (A).
(D) The cranking results corresponding to the cases in (B). \label{F4}}
\end{center}
\end{figure}

Unlike the case in the O(6) limit, the spin-driven QPTs at $\chi\neq0$ cannot be analytically
described because $\sigma$ and $\tau$ are no longer good quantum numbers.
Instead, the cranking method~\cite{Schaaser1986} can be employed to examine the rotational effects on the IBM system~\cite{Alonso1996}.
This method was shown to work well in revealing the shape phase diagram of the cranking IBM~\cite{Cejnar2003} and in identifying
the Jacobi-type transition in deformed nuclei~\cite{Zhang2021}. Particularly, the cranked IBM Hamiltonian has been successfully applied to
give a qualitative estimation of the backbending frequencies in the $A\sim100$ region.
Using the Hamiltonian (\ref{Hp}), we firstly calculate the free energy function of the cranking system,
\begin{eqnarray}\label{freeE}
F(N,\omega)=\langle N, a\mid(\hat{H}^\prime-\omega\hat{L}_x)\mid N, a\rangle\, ,
\end{eqnarray}
where the intrinsic state function is defined by~\cite{Schaaser1986}
\begin{eqnarray}
\mid N, a\rangle=\frac{1}{\sqrt{N!(1+\sum_u\mid a_u\mid^2)^N}}\Big(s^\dag+\sum_ua_ud_u^\dag\Big)^N\mid0\rangle\,
\end{eqnarray}
with $a_u$ ($u=-2,-1,\cdots,2$) denoting the quadrupole parameters. In (\ref{freeE}), $\omega$ represents the cranking frequency, and $\hat{L}_x$ denotes the $x$-component of the angular momentum operator with
\begin{eqnarray}
\hat{L}_x=\frac{1}{\sqrt{2}}(\hat{L}_{-1}-\hat{L}_{+1}),~~\hat{L}_u=\sqrt{10}[d^\dag\times\tilde{d}]_u^{(1)}\, .
\end{eqnarray}
By cranking the system around the x-axis, the symmetries $R_x(\pi)$
and $TR_y(\pi)$ further require
$a_u=a_{-u}$ and $a_u=a_u^*$~\cite{Schaaser1986},
which implies that only $a_0$, $a_1$ and $a_2$ are needed to be determined to
derive the free energy function. Under this constraint, the expectation value of $\hat{L}_x$ can be expressed as~\cite{Schaaser1986}
\begin{eqnarray}
\langle N,a\mid\hat{L}_x\mid N,a\rangle=\frac{2Na_1(\sqrt{6}a_0+2a_2)}{1+\sum_u\mid a_u\mid^2}\, .
\end{eqnarray}
In contrast, the expression of $\langle\hat{H}^\prime\rangle$ is rather complicated even at leading order~\cite{Alonso1996}. Specifically, it is given by
\begin{eqnarray} \nonumber
&~&\langle N,a\mid\hat{H}^\prime(A,B_1,B_2,C,\chi)\mid N,a\rangle\\  \nonumber
&=&\frac{AN(N-1)}{Z}\Big[4(a_0^2+2a_2^2+2a_0^2a_1^2+4a_1^2a_2^2)\\  \nonumber
&+&\chi\sqrt{\frac{32}{7}}(6a_0a_2^2+\sqrt{6}a_1^2a_2-a_0a_1^2-a_0^3)\\  \nonumber
&+&\frac{8\chi^2}{7}\Big(\frac{a_0^4}{4}+a_1^4+a_2^4+a_0^2(\frac{a_1^2}{2}+a_2^2)+a_1^2a_2(\sqrt{6}a_0-a_2)\Big)\Big]\\ \nonumber
&+&\frac{B_1N(N-1)(N-2)(N-3)}{Z^2}64a_1^4(a_0^2+2a_2^2)^2\\ \nonumber
&+&\frac{B_2N(N-1)}{Z}8a_1^2(a_0^2+2a_2^2)\\
&+&\frac{CN(N-1)}{Z}8a_1^2(\sqrt{3}a_0+\sqrt{2}a_2)^2+\cdots\,
\end{eqnarray}
where $Z=(1+\sum_u\mid a_u\mid^2)^2$, and "$\cdots$" denotes contributions from the next leading order terms, which will be
ignored in the concrete calculations. By minimizing the free energy function $F(N,\omega)$ with respect to $a_u$, one can get the optimal values $\bar{a}_u$ for each given $\omega$,
by which the average angular momentum projection $L_x=\langle N,\bar{a}\mid\hat{L}_x\mid N,\bar{a}\rangle$ can be evaluated.
Consequently, the rotational effects on the system can be examined by observing the relationship between the cranking frequency $\omega$ and the angular momentum projection $L_x$.

In FIG.~\ref{F4}, we examine the correlation between $L_x$ and $\omega$ for different $B_1$ values by setting $\chi=-0.3$ and $\chi=-1.32$.
For comparison, the evolution of $\rho(L)$ in the corresponding cases are also analyzed. As shown in FIG.~\ref{F4}, both $B_1$ and $\chi$ can influence the critical behavior of a rotating IBM system. Specifically, the results for $\rho(L)$ indicate that the critical point shifts to higher angular momentum as $|B_1|$ decrease. Meanwhile, the QPT signature weakens and eventually disappears when $|B_1|<0.00017$. This phenomenon can be partially explained from Eq.~(\ref{tc}), which suggests that the spin-driven QPT in the O(6) DS will disappear when $|B_1|<B_2/(2N^2+6N)$.
The transitional processes can be alternatively understood from the cranking results given in FIG.~\ref{F4}(C)-(D). Generally, $L_x$ increase monotonically with the cranking frequency $\omega$. Nevertheless, for cases where $|B_1|\geq0.00017$, $L_x$ may abruptly jump from $L_x\sim12-16$ to $L_x\sim24$ and then remain constant at $L_x=24$ as $\omega$ continues to increase. Here, $L_x=24$ represents the highest angular momentum for $N=12$. Such a dramatic change in $L_x$ can be regarded as a signal of the QPT in the cranking system. The resulting critical angular momenta $L_x\sim12-16$, yielding $L\sim11.5-15.5$ based on the approximations $L_x\approx \sqrt{L(L+1)}$~\cite{Regan2003}, are generally consistent with those predicted from the quantal calculations for $\rho(L)$ in the same cases. Additionally, if the parameters are confined to $|B_1|<0.00017$, $L_x$ will continuously increase with $\omega$ until it reaches $L_x=24$. The results agree well with with the quantal calculations shown in FIG.~\ref{F4}(A)-(B), where the order parameter $\rho(L)$ smoothly increase until $\rho=1.0$ when $|B_1|<0.00017$, indicating that no QPT occurs in these cases.

\begin{figure}
\begin{center}
\includegraphics[scale=0.25]{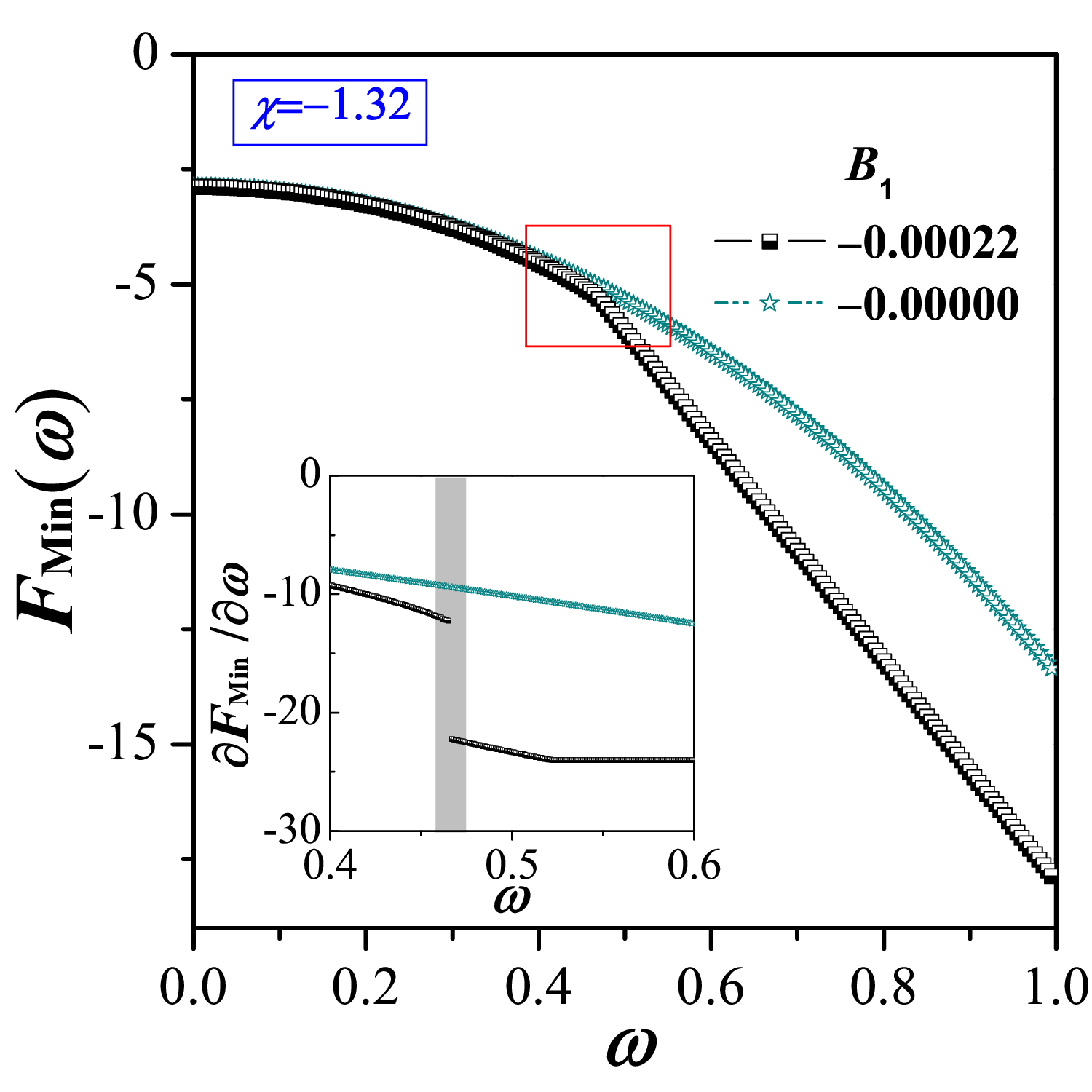}
\caption{The minimal values of the free energy function, $F_\mathrm{Min}(\omega)$, obtained from the two cases with $(\chi,B_1)=(-1.32,-0.00022)$ and $(-1.32,-0.00000)$ in FIG.~\ref{F4}(D), are shown
as functions of $\omega$. The inset highlights the first derivative $\frac{\partial F_\mathrm{Min}}{\partial \omega}$. \label{F42}}
\end{center}
\end{figure}

In fact, the nature of the QPT can be directly understood from the evolutions of the free energy function, $F(N,\omega)$. To illustrate this, we present an example corresponding to the two cases ($B_1=-0.00022$ and $B_1=0.0$) shown in FIG.~\ref{F4}(D). As clearly observed from FIG.~\ref{F42}, the minimal values of the free energy, $F_\mathrm{Min}$, vary continuously as a function of $\omega$ in both cases. However, the first-order derivatives, $\frac{\partial F_\mathrm{Min}}{\partial \omega}$, exhibit a discontinuous change near $\omega\sim0.46$ for $B_1=-0.00022$, thereby suggesting a first-order QPT in this case. This observation is consistent with the understanding of the QPT from order parameter.

Based on the classical formula $\Gamma_x=L_x/\omega$~\cite{Bohrbook,Alonso1996}, we can extract the moments of inertia $\Gamma_x$ from the $L_x-\omega_x$ correlation shown in FIG.~\ref{F4}(C)-(D). For instance, for $(\chi,~B_1)=(-0.3,~-0.00022)$, the moment of inertia initially increase from $\Gamma_x\approx19$ to $\Gamma_x\approx25$ in the range $L_x\in[0-12]$ and then jump to $\Gamma_x\approx50$ near $L_x=24$. These results are also consistent with the quantal calculation for the effective moment of inertia $\Gamma(L)$ defined earlier. Specifically, The results in the same case give $\Gamma\approx13-24$ in the range of $L\in[0-10]$ and $\Gamma\approx45-50$ for $L\in[12-24]$. An example for $\chi=0$ has already been provided in FIG.~\ref{F2}(c). It should be noted that the cranking system with a given $\omega$ produces only one $L_x$ value through the variation calculation. Therefore, a skip in $L_x$, as observed from FIG.~\ref{F4}(C), suggests that the rotating system like that for $B_1=-0.00022$ corresponds to a smaller $\omega$ when $L_x>L_\mathrm{c}$. It means that the $L_x-\omega_x$ correlation in such cases cannot be resolved using the cranking calculations. However, the frequency $\omega$ can be inversely determined from the effective moment of inertia by assuming $\Gamma(L)\approx\Gamma_x=L_x/\omega$ and $L_x\approx\sqrt{L(L+1)}$. For example, this yields $\omega\sim0.43$ for $L=10$ and $\omega\sim0.33$ for $L=16$ in the case of $B_1=-0.00022$, confirming that the frequency for $L>L_\mathrm{c}$ is indeed smaller. In this case, one can expect an "S" shape evolution in the $L_x-\omega_x$ correlation, indicating rotational alignment~\cite{Regan2003}, to appear in the spin-driven QPT process.

As pointed out by Mottelson and Valatin~\cite{Mottelson1960}, the competition between the rotational effect
and paring correlation may result in a rotation-induced phase transition from the superfluid (superconductor) phase to the normal phase.
During this process, the Coriolis forces tend to decouple the pairing correlations, and the normal phase may emerge once the pairing energy-gap $\Delta$ vanishes at high spins. It is intriguing to draw an analogy between the superfluid to normal rotational phase transition and the presently discussed triaxially to axially rotational QPT, especially since the high-order term in the present model is motivated by the pairing interaction at nucleon level. If we confine the QPT picture to the O(6) limit, it becomes evident that the triaxially rotational phase at low spins corresponds to a boson-pairs dominant case with a small number of unpaired $d$ boson (corresponding to small $\tau$ values). In contrast, the axially rotational phase, characterized by a $d$-boson condensate without $s$ bosons, represents a fully boson-pair decoupled situation. This scenario closely mirrors to the transition from the superfluid phase to the normal phase. Note that two $s$ bosons in the IBM are automatically coupled into a boson pair.
Additionally, a significant enhancement in the moment of inertia around the critical point should be also observed in the superfluid to normal phase transition~\cite{Mottelson1960}. Therefore, from the perspective of collective rotation breaking pairs (boson-type or fermion-type), the presently discussed QPT can be viewed as a superfluid to normal phase transition but manifested in the boson model language.

\begin{figure}
\begin{center}
\includegraphics[scale=0.36]{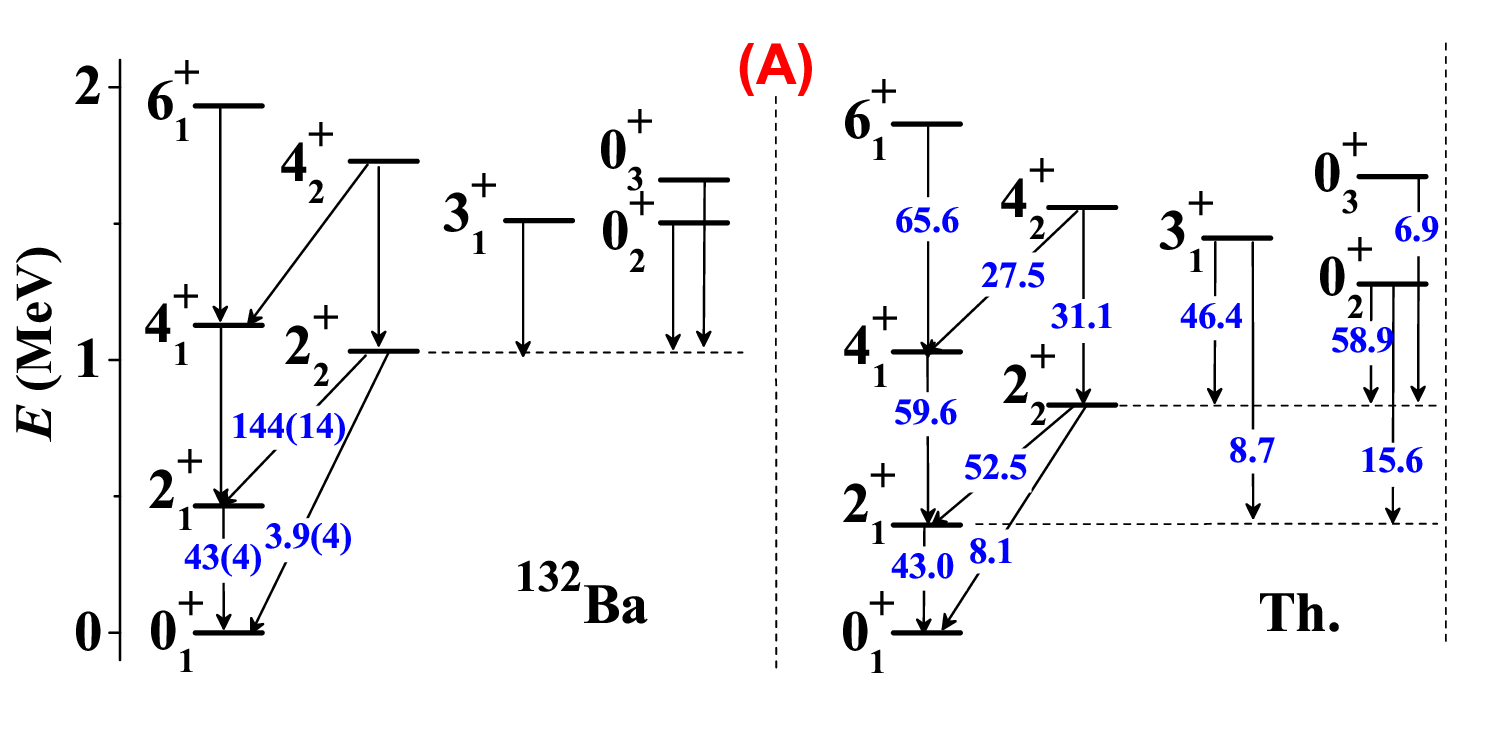}
\includegraphics[scale=0.36]{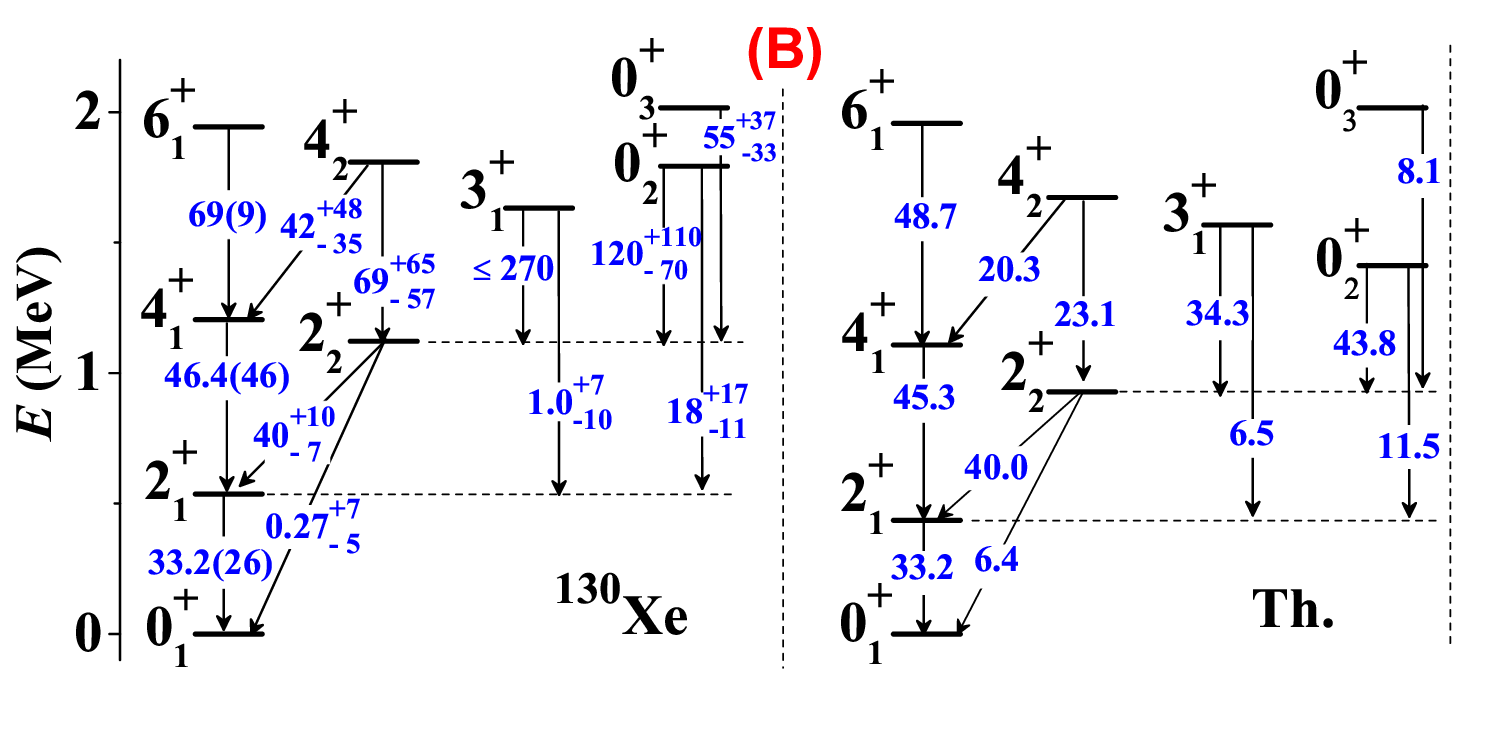}
\caption{The low-lying structures of $^{132}$Ba~\cite{Khazov2005} and $^{130}$Xe~\cite{Singh2001,Peters2016} are compared with
the results obtained from the Hamiltonian (\ref{Hp}) with $\chi=-0.02$ and the following parameters (in
MeV): $A=-0.038(-0.056),~B_1=-0.0005(-0.0008),~B_2=0.080(0.093)$, and
$C=0.014(0.013)$ for $^{132}$Ba ($^{130}$Xe). In the calculations for $B(E2)$ (in W.u.), the effective charge in (\ref{E2}) is set to $e_\alpha=0.076~\mathrm{eb}$ for $^{132}$Ba and $e_\alpha=0.08~\mathrm{eb}$ for $^{130}$Xe, with the dimensionless parameter fixed at $\theta=-1.7$.\label{F5}}
\end{center}
\end{figure}

\begin{figure}
\begin{center}
\includegraphics[scale=0.25]{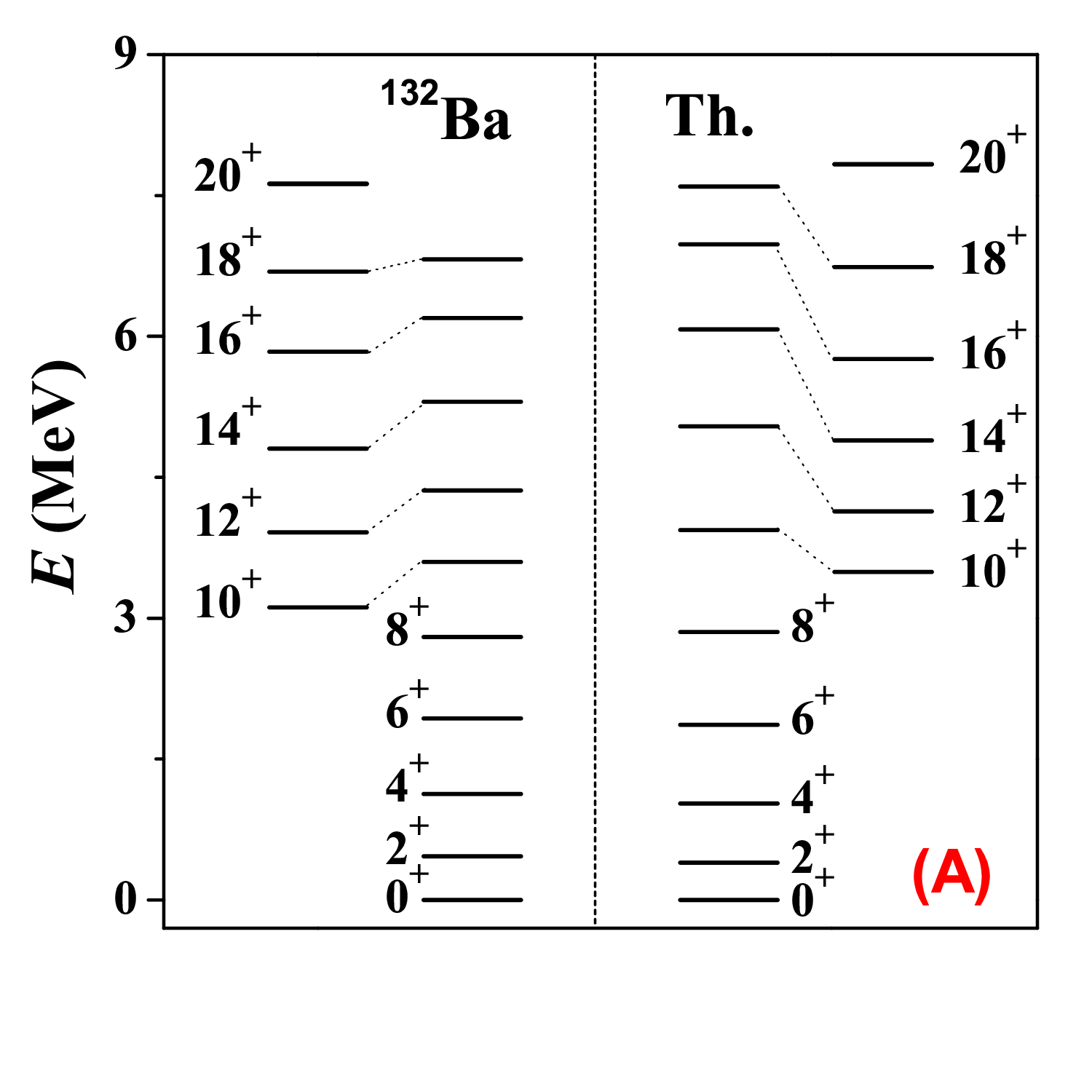}
\includegraphics[scale=0.25]{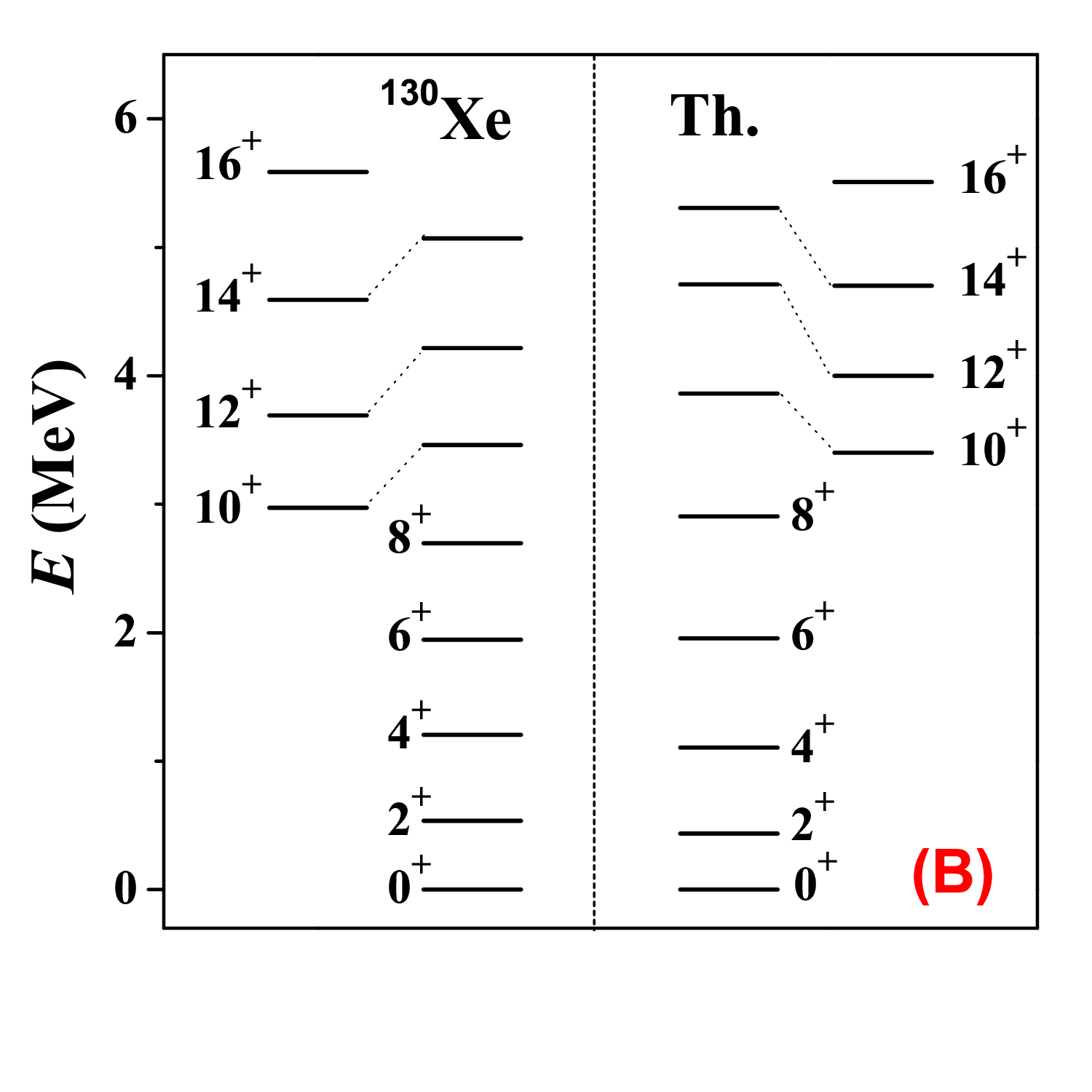}
\caption{(A) The yrast levels and yare levels in $^{132}$Ba are shown, with the same angular momentum $L$ connected by short dashed lines. For comparison,
the theoretical results for the yrast levels as well as those associated with strong $B(E2;L\rightarrow L-2)$ cascades related to $8_1^+$
are presented alongside. (B) The same as in (A), but for $^{130}$Xe.  \label{F6}}
\end{center}
\end{figure}

\begin{figure*}
\begin{center}
\includegraphics[scale=0.18]{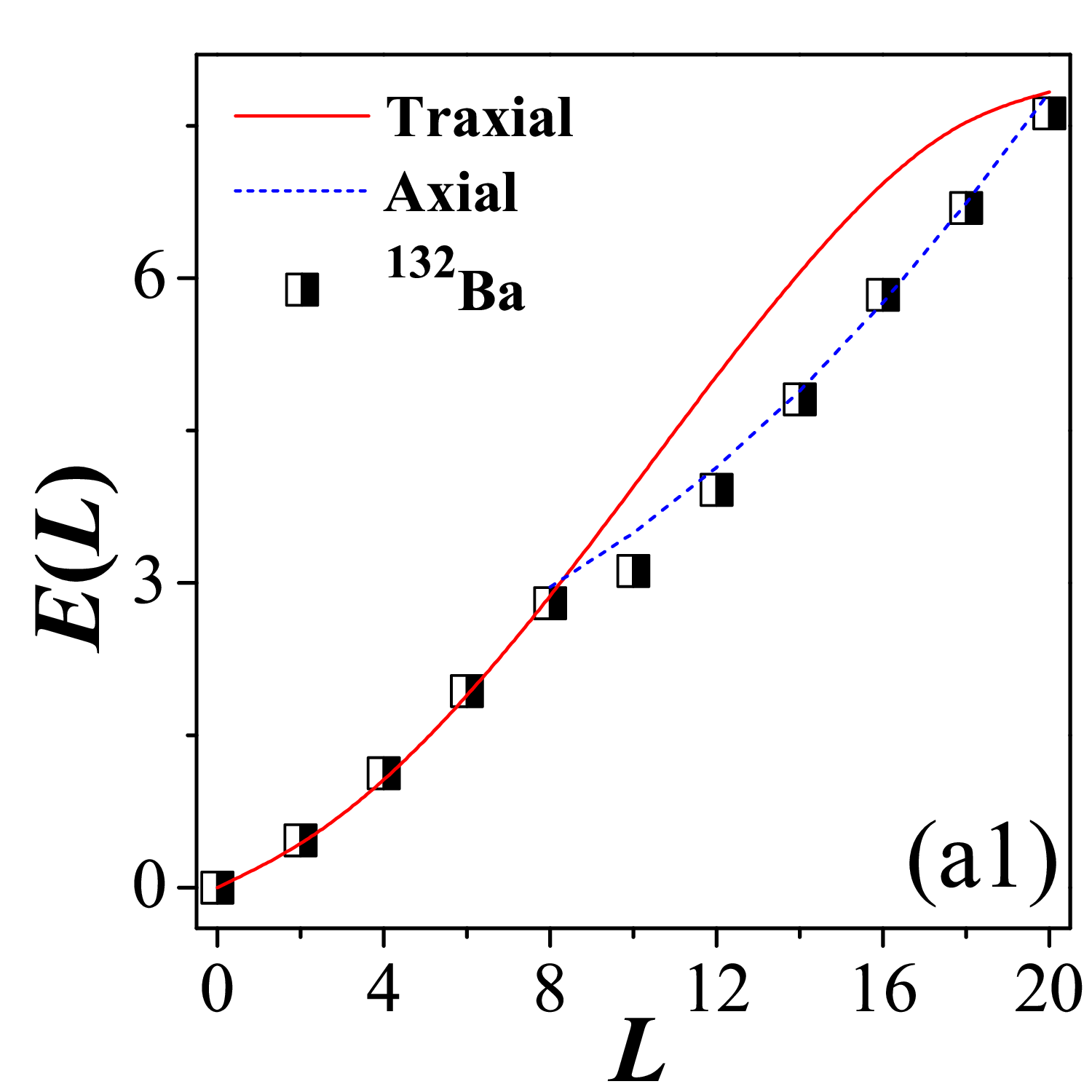}
\includegraphics[scale=0.18]{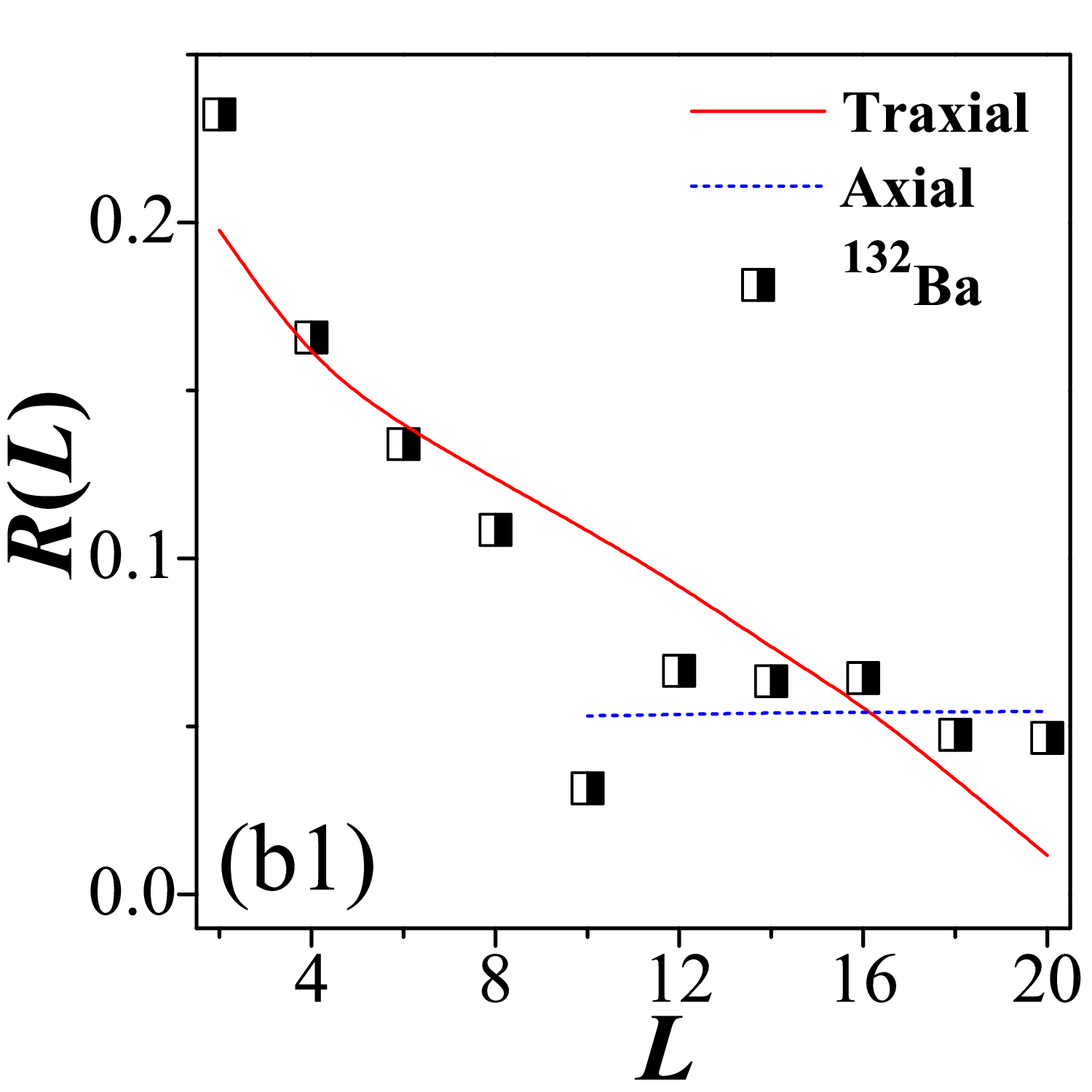}
\includegraphics[scale=0.18]{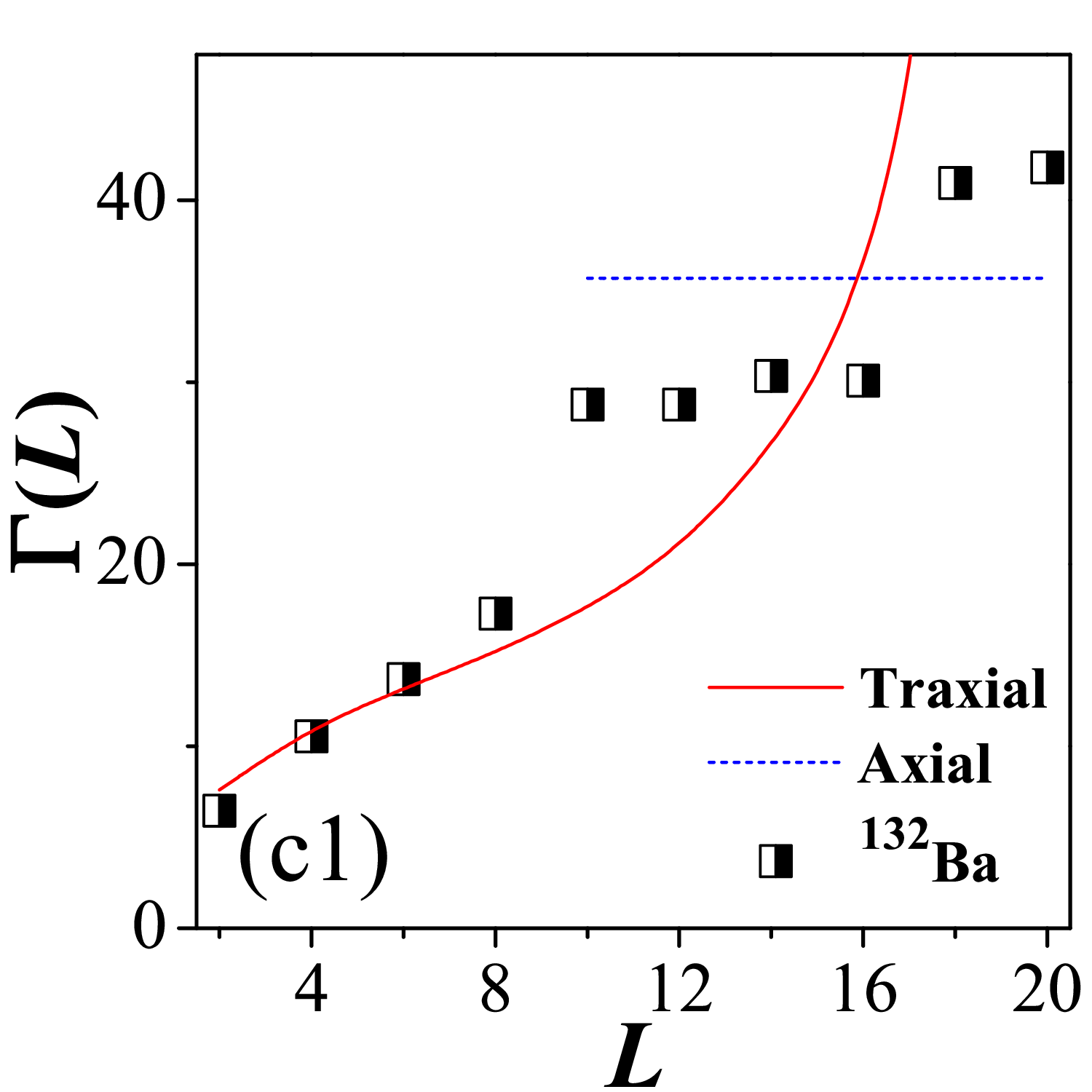}
\includegraphics[scale=0.18]{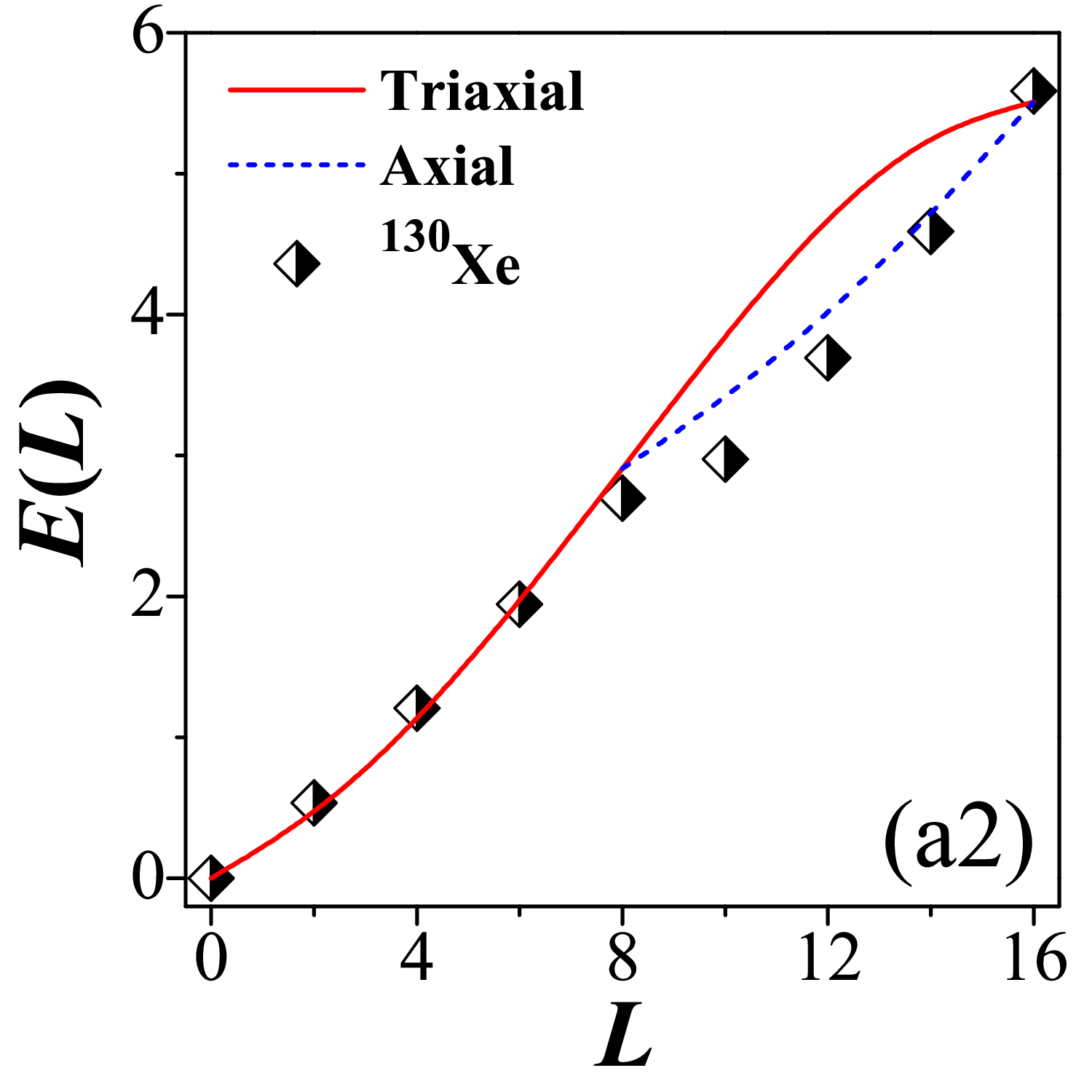}
\includegraphics[scale=0.18]{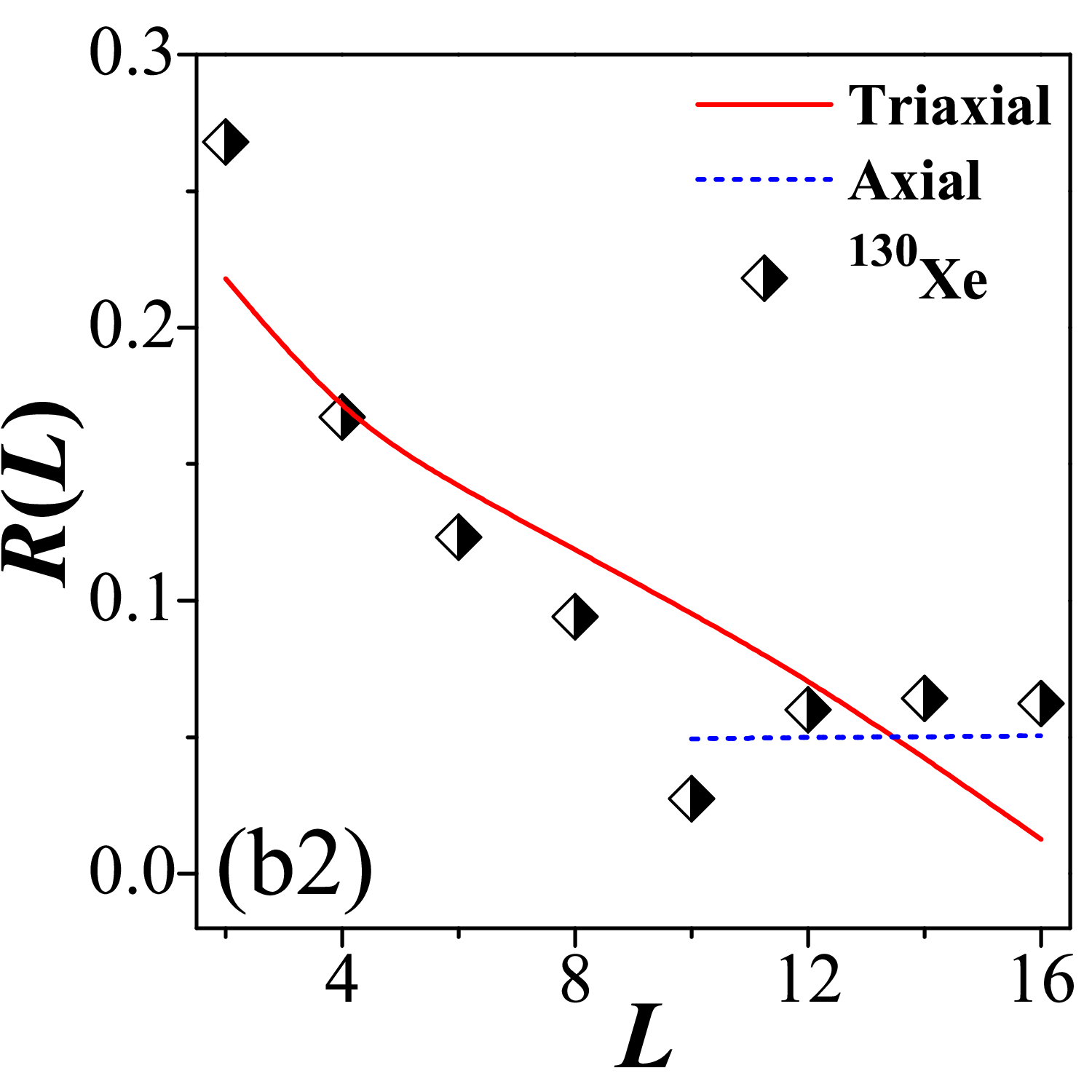}
\includegraphics[scale=0.18]{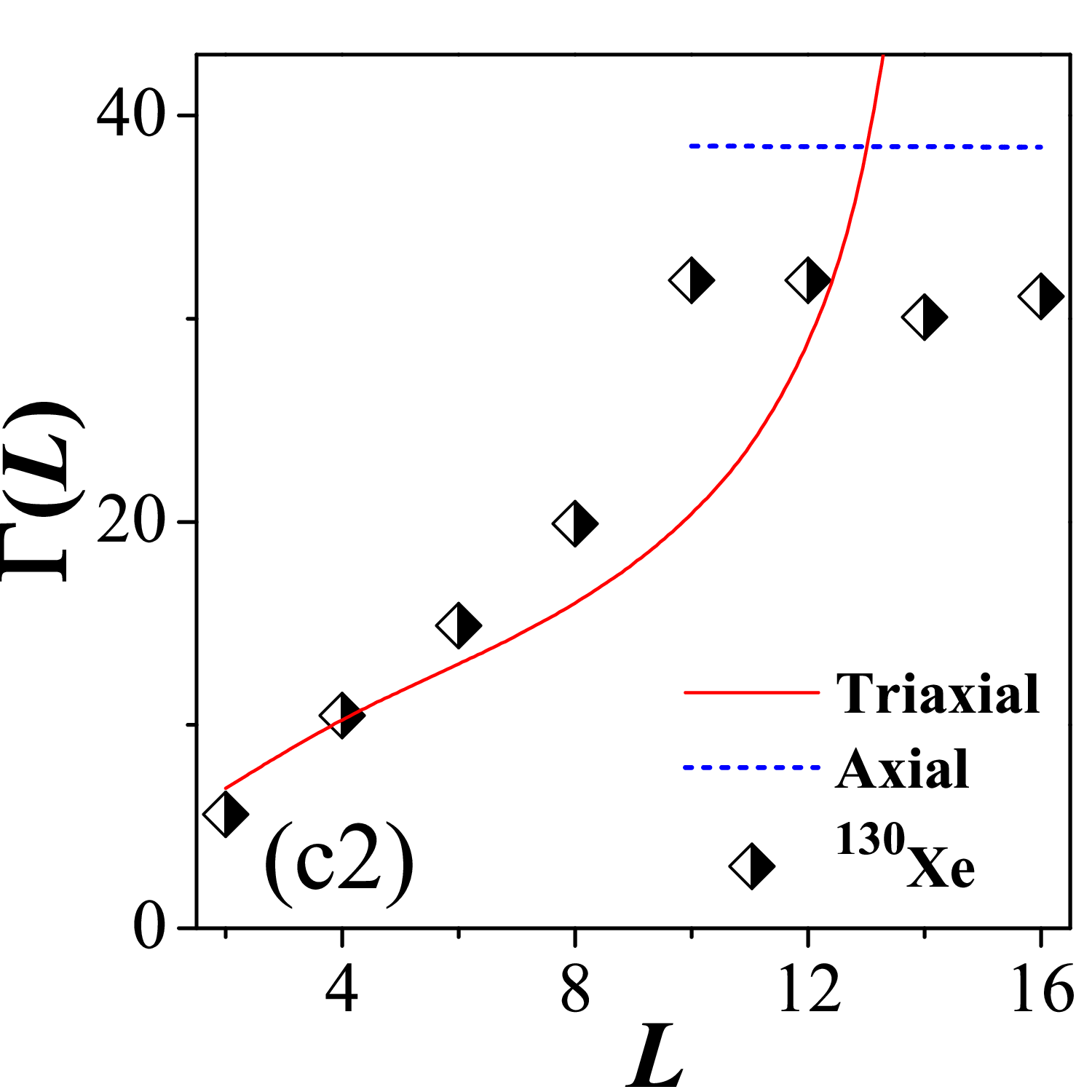}
\caption{The level energy ($E(L)$ in MeV), E-GOS curve ($R(L)$ in MeV$/\hbar$), and the effective moment of inertia ($\Gamma(L)$ in MeV$^{-1}\hbar^2$) for $^{132}$Ba and $^{130}$Xe are extracted from the experimental data for the yrast levels. The theoretical results, denoted as "Triaxial" and "Axial" (corresponding to line 1 and line 2 in FIG.~\ref{F2}), are obtained from the respective columns in the corresponding panels of FIG.~\ref{F6}.\label{F7}}
\end{center}
\end{figure*}

Strictly speaking, a QPT can only be rigorously defined in the classical limit, which implies infinite systems at zero temperature. Nevertheless, the concept of QPT has been successfully applied to
describe structural evolutions in finite nuclei~\cite{CJC2010}, as the main features of a QPT may persist for mesoscopic systems with $N\sim10$, as pointed out by Iachello and Zamfir~\cite{Iachello2004,Iachello2006}. Similar to ground-state QPTs (GSQPTs) in the IBM~\cite{IachelloBook}, the proposed triaxial to axial shape transition is governed by a quantum Hamiltonian and classified using a zero-temperature free energy function. Its occurrence requires that the interaction parameters satisfy specific conditions (see Fig~.\ref{F1}). Therefore, the transitional phenomenon discussed in this work can be interpreted as a QPT, despite not being associated with any changes in ground states. On the other hand, the concept of QPT has been generalized to excited states~\cite{Cejnar2006,CCF2008,Iachello2018,Caprio2019,Cejnar2021}, leading to the so-called excited-state QPT (ESQPTs). A firm theoretical ground for the ESQPTs in many-body systems has been established by Caprio, Cejnar and Iachello in \cite{CCF2008}. Their analysis indicates that ESQPT phenomena may universally occur in a broad family of two-level fermionic and bosonic models~\cite{CCF2008}. Although ESQPTs in the IBM are considered to be rooted in the associated GSQPTs, such as the U(5)-O(6) and U(5)-SU(3) transitions, some similarities can be identified between the presently discussed rotational QPT and a certain type of ESQPT, as both occur in excited states. For example, in the ESQPT associated with the U(5)-O(6) transition, an O(6) like structure is approximately preserved at low energy, with the the $d$-boson occupation probability serving as the order parameter~\cite{CCF2008}. The difference lies in the fact that the ESQPTs typically manifest singularities in the level density of vibrational spectrum as excitation energy increase, whereas the transition discussed here describes shape change in rotational states and thus occurs as a function of angular momentum or rotational frequency. Additionally, it is worth mentioning the Jacobi-type transitions~\cite{Zhang2017}, which are supposed to accompany the ESQPTS associated with the U(5)-SU(3) or O(6)-SU(3) transitions. Interestingly, the Jacobi-type transitions also describe shape changes as a function of angular momentum but manifest as transitions from axial ($\gamma$-rigid) to triaxial ($\gamma$-soft) shapes~\cite{Zhang2017}, yielding a transitional direction opposite to the present triaxial to axial shape transition. Moreover, the Jacobi-type transitions typically occur at lower spins. Nonetheless, all these transitions are established within the algebraic framework, indicating that the IBM serves as an ideal theoretical laboratory for analyzing various QPT phenomena in many-body systems.

\begin{center}
\vskip.2cm\textbf{IV. Test of the model description}
\end{center}\vskip.2cm

As discussed above, the new model suggests that a triaxially to axially rotational QPT may occur in an O(6)-like system.
Experimentally, nuclei near $A=130$ have traditionally been considered to exhibit the O(6) DS~\cite{Casten1985,Brentano1988}. To provide a preliminary test of the model,
we focus on two specific nuclei: $^{132}$Ba~\cite{Khazov2005} and $^{130}$Xe~\cite{Singh2001,Peters2016}. Both nuclei are suggested to possess triaxially deformed ground-state shapes~\cite{Paul1989,Morrison2020} with
low-lying structures close to the O(6) DS~\cite{Kuhn1996,Neuneyer1996}. Additionally, these nuclei exhibit transitional phenomena associated with their yrast states~\cite{Paul1989}, which have been
explained from the point of band crossing based on the alignment of the $h_{11/2}$ neutron holes~\cite{Higashiyama2003,Cheng2015,Lei2020}.
Here, we examine the relevant phenomena from the perspective of the spin-driven QPT, emphasizing changes in the collective mode of a rotating quantum system.
To account for possible corrections due to breaking the O(6) DS, we apply the model Hamiltonian (\ref{Hp}) to analyze the experimental data.
Despite this, the O(6) symmetry language can still be used to interpret the model results, as the O(6) DS is supposed to be lightly broken in the present case.

Traditionally, in the IBM calculations,
the total boson number should correspond to the number of valence nucleons (or holes) pairs for a given nucleus~\cite{IachelloBook}. This result in $N=6$ for $^{132}$Ba and $N=5$ for $^{130}$Xe, corresponding to $L\leq12$ and $L\leq10$, respectively, for the two nuclei. Such a constraint aligns with the traditional applications of the IBM for describing low-lying states in nuclei. However, the present model aims to describe both low-lying properties and high-spin phenomenon in triaxial nuclei. To include high-spin yrast states, the effective total boson number in the model is phenomenally set as $2N=L_\mathrm{max}$, where $L_\mathrm{max}$ represents the maximal angular momentum involved in the high-spin phenomenon. Early IBM calculations~\cite{Scholten1983,Heyde1982,Wolf1982} indicate that an effective boson number may be necessary to capture the spectroscopic properties when the intruder orbits play an important role. Specifically, the effective boson number is set here to $N=10$ for $^{132}$Ba and $N=8$ for $^{130}$Xe.

{\small\begin{table} \caption{The available data~\cite{Morrison2020} for quadrupole moments
(in eb) are compared with the theoretical predictions and the shell model results (SM) taken from in \cite{Kaneko2023}. } \label{T1}
\begin{tabular}{cccc|cccc}\hline\hline
&$^{132}$Ba&Th.&&$^{130}$Xe&Th.&SM\\
\cline{2-7}
$Q(2_1^+)$&-&-0.50&&-0.38$_{-14}^{+17}$&-0.44&-0.26\\
$Q(4_1^+)$&-&-0.76&&-0.41(12)&-0.69&-0.32\\
$Q(2_2^+)$&-&0.22&&0.1(1)&0.20&0.25\\
\hline\hline
\end{tabular}
\end{table}}

To determine the model parameters, the low-energy patterns below 2 MeV in both nuclei are fitted using the model Hamiltonian (\ref{Hp}).  As shown in FIG.~\ref{F5}, the experimental data for both level energies and $B(E2)$ transitions are generally well described by the theoretical results obtained from the model Hamiltonian with a very small $\chi$ value ($\chi=-0.02$). Therefore, it is reasonable to attribute the low-spin excitations below 2 MeV to the O(6)-like triaxial modes. Specifically, the patterns in FIG.~\ref{F5} can be approximately understood within of the O(6) IRREPs ($\chi=0$) with $\sigma=N$ and $\tau=0,~1,~2,~3$, while
two $0^+$ excitations may be attributed to mixing between $(\sigma,~\tau)=(N,~3)$ and $(N-2,~0)$. Above 2 MeV, collective states in both nuclei become difficult to classify within an O(6)-like pattern. Instead, the high-spin states can be recognized as members of rotational bands.
In addition, numerical calculations indicate that adopting a relatively larger boson number can improve the O(6) description of the $B(E2)$ transitions in experiments. For instance, increasing the boson number from $N=5$ to $N=8$ may enhance the ratio $B(E2;6_1^+\rightarrow4_1^+)/B(E2;2_1^+\rightarrow0_1^+)$ by approximately $20\%$. Furthermore, as seen in Table~\ref{T1}, the predicted quadrupole moments agree well with the available data~\cite{Morrison2020} and the shell model calculations~\cite{Kaneko2023}. Particularly, a sign change from $Q(2_1^+)$ to $Q(2_2^+)$, indicating $\gamma$ instability, has been well reproduced by the theoretical calculations. These results suggest that a relative large $|\theta|$ value should be adopted in order to accurately reproduce the $E2$ properties of the O(6)-like nuclei. It should be emphasized that in calculating the $B(E2)$ transitions, we do not adopt a consistent-$Q$ formalism. Specifically, the dimensionless parameter $\theta$ in the $E2$ operator (\ref{E2}) takes a value different from that of $\chi$, which is used in the Hamiltonian (\ref{Hp}).

To test the model description of high-spin properties, the yrast sequences up to $L_\mathrm{max}$ in both nuclei
are shown in FIG.~\ref{F6}. It can be observed that the experimental level energies can be well reproduced by the theoretical calculations up to $L=20$ for $^{132}$Ba and $L=16$ for $^{130}$Xe.
The yrast levels have been classified into two parts (bands), forming a picture of bands crossing. Note that the yrare states ($10_2^+,12_2^+,\cdots$) listed in experiments serve only as a reference for bands crossing and
cannot be directly assigned to members of the ground-state bands due to the lack of $B(E2)$ data for these states. On the theoretical side, the yrast levels can be approximately separated according to the O(5) IRREPs: the low-spin levels with $L\leq8$ correspond to $\tau=L/2$, while the high-spin levels with $L\geq10$ correspond to $\tau=N$. A sudden diminishment in the energy gaps between adjacent yrast levels is observed near $E(8_1^+)$, which is considered as a preliminary signature of shape change in a rotating nuclear system.

To discern different collective modes along the yrast line, we further examine the evolutions of $E(L)$, $R(L)$ and $\Gamma(L)$. As shown in FIG.~\ref{F7}, the experimental evolutional features of these quantities are well reproduced by the model calculations, especially for the discontinuous changes appearing around $L=8$. This suggests that the yrast state evolutions in both $^{132}$Ba and $^{130}$Xe can be integrated as a triaxially to axially rotational QPT with the critical angular momentum $L_\mathrm{c}=8$. This interpretation is analogous to the prescription for structural evolutions along the yrast lines in spherical nuclei provided by Regan {\it et~al.}~\cite{Regan2003}.
In fact, early analysis of $^{132}$Ba~\cite{Paul1989} suggested that this nucleus exhibits maximal triaxiality with $\gamma\approx-30^\circ$ at low spins but change to an oblate deformation with $\gamma\approx-60^\circ$ at high spins. Such a shape change from triaxial to axial is consistent with the present triaxially to axially rotational QPT explanation. In contrast, the band-crossing description along cannot account for how the collective mode (or shape) changes during the yrast state evolutions. Additionally, one may observe from FIG.~\ref{F7}(c1) that there exists a second enhancement in $\Gamma(L)$ near $L=18$ for $^{132}$Ba, which is not well captured in the current scheme.

{\small\begin{table*} \caption{The available experimental data~\cite{Khazov2005,Morrison2020} for the $B(E2;L\rightarrow L-2)$ transitions
(in W.u.) along the yrast line are compared with the results predicted by the model calculations. "-" indicates that the value is unknown experimentally} \label{T2}
\begin{tabular}{ccccccccccccc}\hline\hline
$L_i^\pi\rightarrow
L_f^\pi$&$2_1^+\rightarrow0_1^+$&$4_1^+\rightarrow2_1^+$&$6_1^+\rightarrow4_1^+$&
$8_1^+\rightarrow6_1^+$&$10_1^+\rightarrow8_1^+$&$12_1^+\rightarrow10_1^+$&
$14_1^+\rightarrow12_1^+$&$16_1^+\rightarrow14_1^+$&$18_1^+\rightarrow16_1^+$&$20_1^+\rightarrow18_1^+$\\
\hline
Th.&43&59.6&65.6&66.4&0.00&10.5&14.7&17.4&16.9&11.5\\
$^{132}$Ba&43(4)&-&-&-&0.46(1)&-&-&-&-&-\\  \hline
Th.&33.2&45.3&48.7&46.0&0.24&12.0&13.8&10.4&&\\
$^{130}$Xe&33.2(26)&46.4(46)&69(9)&-&1.69(4)&-&-&-&&\\
\hline\hline
\end{tabular}
\end{table*}}

As analyzed above, a rotational phase transition may occur in both nuclei at $L_\mathrm{c}=8$, which imposes a stringent constraint on the model parameters as seen from Eq.~(\ref{tc}). Using the classical formula $\omega=\frac{\partial E}{\partial L}\approx\frac{E_\gamma(L\rightarrow L-2)}{2}$, the critical frequency $\omega_\mathrm{c}$ (in $\mathrm{MeV}/\hbar$) can be extracted from the yrast spectra. Specifically, the data indicate $\omega_\mathrm{c}\approx0.43$ for $^{132}$Ba and $\omega_\mathrm{c}\approx0.38$ for $^{130}$Xe. The corresponding theoretical values are $\omega_\mathrm{c}=0.49$ and $\omega_\mathrm{c}=0.47$, respectively. Alternatively, one can get the critical frequency $\omega_\mathrm{c}$ from the cranking calculations, yielding $\omega_\mathrm{c}\approx0.58$ for $^{132}$Ba and $\omega_\mathrm{c}\approx0.51$ for $^{130}$Xe. This suggests that the experimental critical frequencies can be roughly estimated from the semi-classical analysis of the model. A complementary analysis of the QPT can be provided by examining the $B(E2)$ transitions, as observed from TABLE~\ref{T2}. Although limited data available, the rather weak transitional strength with $B(E2;10_1^+\rightarrow8_1^+)/B(E2;2_1^+\rightarrow0_1^+)<0.06$ for both nuclei provides
another key signature of the QPT occurring at $L_\mathrm{c}=8$. Additionally, the predicted $B(E2)$ results suggest that high-spin states still maintain collective band structures.

\begin{center}
\vskip.2cm\textbf{V. Summary}
\end{center}\vskip.2cm

In summary, an algebraic approach to describe the spin-driven QPT in triaxial nuclei has been proposed in the IBM frame through introducing a high-order interactional term in the O(6) DS. The resulting model can yield a first-order transition along the yrast line, characterized by a significant enhancement in the moment of inertia. Different aspects of the spin-driven QPT
have been analyzed in detail, with an analogy to the superfluid-normal phase transition and a comparison with the excited-state QPT being emphasized. In particular, it is shown that the QPT features can be well preserved even in systems deviating away from the O(6) limit. As a preliminary test, we have applied this algebraic scheme to explain the transitional phenomena along the yrast lines in $^{132}$Ba and $^{130}$Xe. The results confirm that the proposed model can provide a simple yet effective way to simultaneously describe the yrast state evolutions and low-lying dynamics in O(6)-like nuclei.  It is worth noting that another method of generating high spin states in the IBM involves introducing two unpaired fermions~\cite{Gelberg1980,Yoshida1982}, by which high-spin states are explained as members of a rotational band built on two quasi-particle excitation.
In contrast, high-spin states in the present scheme are generated simply by the $d$-boson condensate configuration, which may become energetically favourable due to the high-order correction motivated from pairing correlations.
However, whether this approach can be extended to describe high-spin phenomena in other systems, such as those in well-deformed nuclei, requires further investigations. Related work is in progress.

\bigskip

\begin{acknowledgments}
Support from the National Natural Science Foundation of China
(12375113, 12247107, 12175097), the National Key Basic Research Program of China (2015CB856900)
and from LSU through its Sponsored Research Rebate Program as well as the LSU Foundation's Distinguished Research Professorship Program is acknowledged.
\end{acknowledgments}

\end{document}